\documentclass{elsart}
\usepackage{amsmath}
\usepackage{amsfonts}
\usepackage{graphics}
\usepackage{subfigure}
\usepackage{epsfig}
\usepackage{rotate}
\usepackage{makeidx}
\usepackage{graphicx}

\begin{document}

\begin{frontmatter}
\title{Application of the GALI Method to Localization Dynamics in Nonlinear Systems}

\author{}
T. Bountis$^{\mathrm{a}}$, \ead{bountis@math.upatras.gr}
\ead[url]{http://www.math.upatras.gr/~bountis/} T.
Manos$^{\mathrm{a},\mathrm{b}}$ and
H.~Christodoulidi$^{\mathrm{a}}$\\[12pt]

\small\itshape
$^a$Center for Research and Applications of Nonlinear
Systems (CRANS), Department of Mathematics, University of Patras,
GR--26500, Patras, Greece.\\
$^b$Laboratoire d'Astrophysique de Marseille (LAM), Observatoire
Astronomique de Marseille-Provence (OAMP), 2 Place Le Verrier,
F--13248, Marseille, France.

%=============================================
\begin{abstract}
We investigate localization phenomena and stability properties of
quasiperiodic oscillations in $N$ degree of freedom Hamiltonian
systems and $N$ coupled symplectic maps. In particular, we study an
example of a parametrically driven Hamiltonian lattice with only
quartic coupling terms and a system of $N$ coupled standard maps. We
explore their dynamics using the Generalized Alignment Index (GALI),
which constitutes a recently developed numerical method for
detecting chaotic orbits in many dimensions, estimating the
dimensionality of quasiperiodic tori and predicting slow diffusion
in a way that is faster and more reliable than many other approaches
known to date.
\end{abstract}
%=============================================
\vspace{0.15cm}

We dedicate this paper to Professor Apostolos Hadjidimos, whose
inspiring research and teaching all these years has demonstrated
that Numerical Analysis is a truly fundamental branch of Mathematics
and not simply a useful tool for solving scientific problems.
\end{frontmatter}
\vspace{-0.5cm}
%=============================================
\section{Introduction}\vspace{-0.75cm}
\label{intro}

In the present paper, we apply the Generalized Alignment Index
(GALI) method introduced by Skokos et al (2007) \cite{sk:6} to two
types of dynamics, which are related to localization  phenomena in
multi--dimensional Hamiltonian systems and symplectic maps. The
first type refers to quasiperiodic oscillations in the vicinity of
what is known as \textit{discrete breathers}, or exact periodic
solutions of multi--dimensional systems, which are exponentially
localized in real space. The second class of phenomena concerns
localization in Fourier space and is evidenced by the persistence
of quasiperiodic recurrences in the neighborhood of normal mode
oscillations of nonlinear lattices, observed for example in the
famous numerical experiments performed by Fermi Pasta and Ulam
(FPU) in the early 1950's \cite{ber,flach1,flach2,antono2}.

In section \ref{GALI_def}, we present a brief introduction to the
GALI method and in section \ref{DB} we use it to study quasiperiodic
motion on tori of low dimensionality near a stable discrete breather
of a Hamiltonian lattice, with nonlinear on site potential and only
quartic nearest neighbor interactions. Then in section \ref{N_sm},
we turn to a system of $N$ coupled standard maps and give evidence
for the existence of both of the above types of localization: First
starting with initial conditions localized in real space, we find
low dimensional quasiperiodic motion, which persists for very long
times. We also study in this discrete model recurrences of its
(linear) normal mode oscillations and find that, in contrast with
the FPU model, the tori associated with them are high--dimensional,
which may be due to the fact that, unlike the FPU example, each map
contains on site nonlinear terms, which depend on its individual
variables. Finally we discuss our conclusions in section
\ref{concl}.

\vspace{-0.5cm}
\section{Definition of the GALI method}\vspace{-0.75cm}
\label{GALI_def}

Let us consider the $2N$--dimensional phase space of a conservative
dynamical system, which may be represented by a Hamiltonian flow of
$N$ degrees of freedom or a $2N$--dimensional system of coupled
symplectic maps. In order to study whether an orbit is chaotic or
not, we examine the asymptotic behavior of $k$ initially linearly
independent deviations from this orbit, denoted by the vectors
$\overrightarrow{\nu}_{1},\overrightarrow{\nu}_{2},...,
\overrightarrow{\nu}_{k}$ with $2\leq k \leq 2N$. Thus, we follow
the orbit, using Hamilton's equations (or the map equations) of
motion and solve in parallel the variational equations about this
orbit to study the behavior of solutions located in its
neighborhood.

The Generalized Alignment Index of order $k$ is a generalization of
the Smaller Alignment Index (SALI) introduced in \cite{sk:1} and is
defined as the norm of the wedge (or exterior) product of $k$
associated unit deviation vectors \cite{sk:6}:\vspace{-0.25cm}
\begin{equation}\label{GALI:0}
    GALI_{k}(t)=\parallel \hat{\nu}_{1}(t)\wedge \hat{\nu}_{2}(t)
    \wedge ... \wedge \hat{\nu}_{k}(t) \parallel \vspace{-0.25cm}
\end{equation}
representing \textit{the volume of the parallelepiped}, whose edges
are these $k$ vectors. We note that the hat ($\,\hat{}\,$) over a
vector denotes that it is of unit magnitude and that $t$ represents
the continuous or discrete time.

In the case of a chaotic orbit, all deviation vectors tend to become
\textit{linearly dependent}, aligning in the direction of the
eigenvector corresponding to the maximal Lyapunov exponent and
GALI$_{k}$ tends exponentially to zero following the law
\cite{sk:6}:
\begin{equation}\label{GALI:1}
GALI_{k}(t)\propto
e^{-[(\sigma_{1}-\sigma_{2})+(\sigma_{1}-\sigma_{3})+...+(\sigma_{1}-\sigma_{k})]t},\vspace{-0.25cm}
\end{equation}
where $\sigma_1> \ldots >\sigma_k$ are approximations of the first
$k$ largest Lyapunov exponents of the dynamics. In the case of
regular motion, on the other hand, all deviation vectors tend to
fall on the $N$--dimensional tangent space of the torus, where the
motion is quasiperiodic. Thus, if we start with $k\leq N$ general
deviation vectors, these will remain \textit{linearly independent}
on the $N$--dimensional tangent space of the torus, since there is
no particular reason for them to become aligned. As a consequence,
GALI$_{k}$ in this case remains practically constant for $k\leq N$.
On the other hand, for $k>N$, GALI$_{k}$ tends to zero, since some
deviation vectors will eventually become \textit{linearly
dependent}, following power laws that depend on the dimensionality
of the torus. According to our asymptotic analysis of determinants
entering in an expansion of (\ref{GALI:0}) in a basis of
eigenvectors following the motion, one obtains the following formula
for the GALI$_{k}$, associated with quasiperiodic orbits lying on
$m$-dimensional tori \cite{ChrisBou}: \vspace{-0.35cm}
\begin{equation}\label{GALI:2}
    GALI_{k}(n)\propto \left\{
                         \begin{array}{ll}
                            constant, & \hbox{if $2\leq k\leq m$}\\
                           \frac{1}{t^{k-m}}, & \hbox{if $m<k\leq 2N-m$} \\
                           \frac{1}{t^{2(k-N)}}, & \hbox{if $2N-m<k\leq 2N$}
                         \end{array} .
                       \right.\vspace{-0.35cm}
\end{equation}
The computation of such determinants, however, is hardly the most
efficient way to compute the GALI$_{k}$, especially in cases where
the dimension of the system $N$ becomes large. For this reason, we
have recently introduced and employed a method based on Singular
Value Decomposition to show that the GALI$_{k}$ can be very
efficiently computed as the product of the singular values of a
$2N\times k$ matrix \cite{LDI}.

\vspace{-0.5cm}
\section{Applications to the dynamics near discrete
breathers}\vspace{-0.75cm} \label{DB}

In this section we examine the dynamics in the vicinity of discrete
breathers in a one--dimensional Hamiltonian lattice with quartic on
site potential and linear dispersion terms in its nearest neighbor
particle interactions. For this purpose, we start with initial
conditions near the exact breather and check whether the motion
remains quasiperiodic or becomes chaotic. We accomplish this by
computing the GALI indices along the reference orbit. If the
breather is stable, the GALI method can be used to determine the
dimensionality of the tori surrounding the breather in the
$2N$--dimensional phase space. This dimension is generically $N$ and
equals the number of frequencies that are being excited in the
neighborhood of the exact breather. As for chaotic motion diffusing
slowly away from these breathers, it is rapidly and efficiently
predicted by the exponential convergence of all GALI indices to
zero.
%=============================================
\begin{figure*}[h]
  \includegraphics[width=8.5cm]{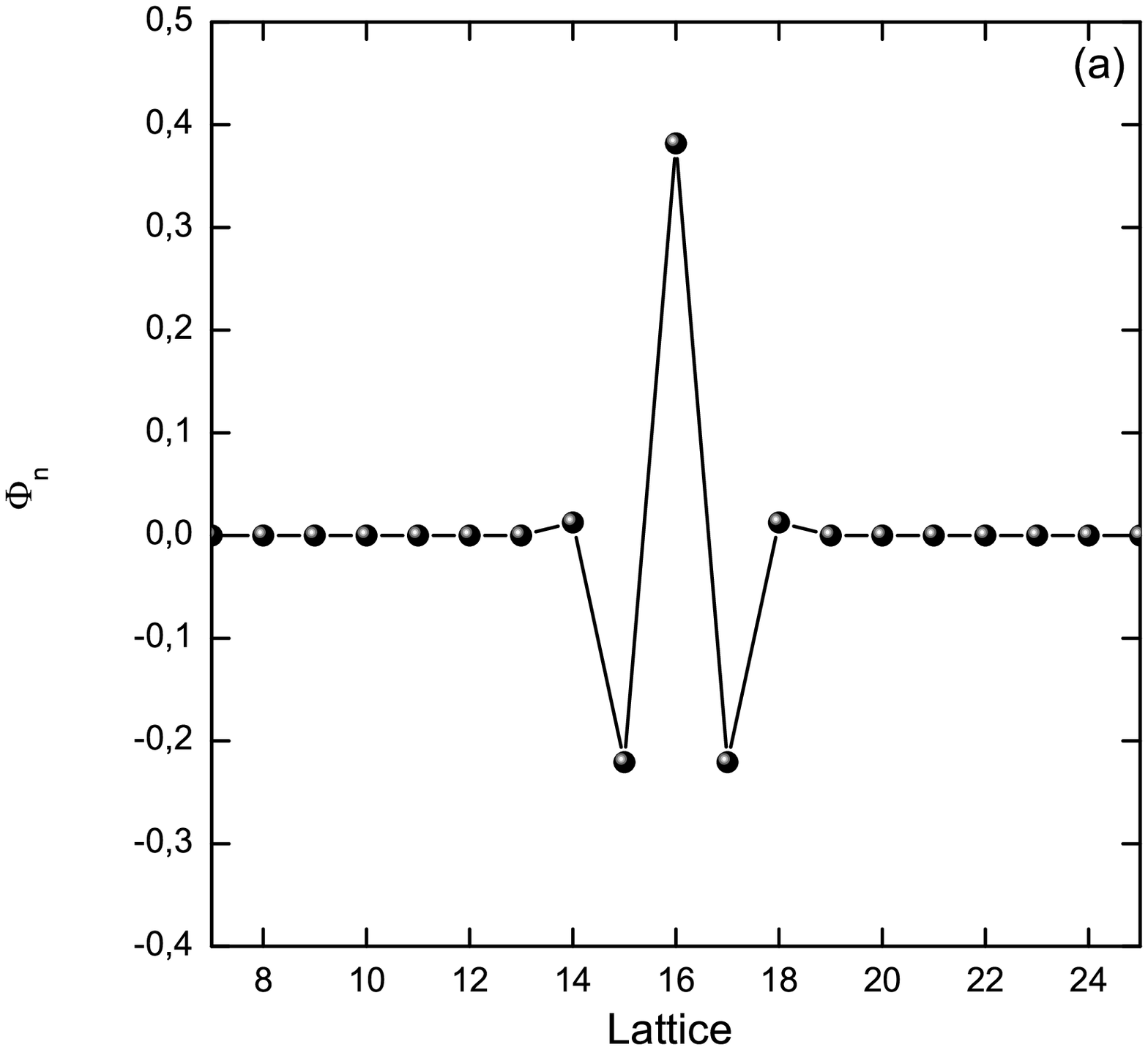}\hspace{-1.5cm}
  \includegraphics[width=8.5cm]{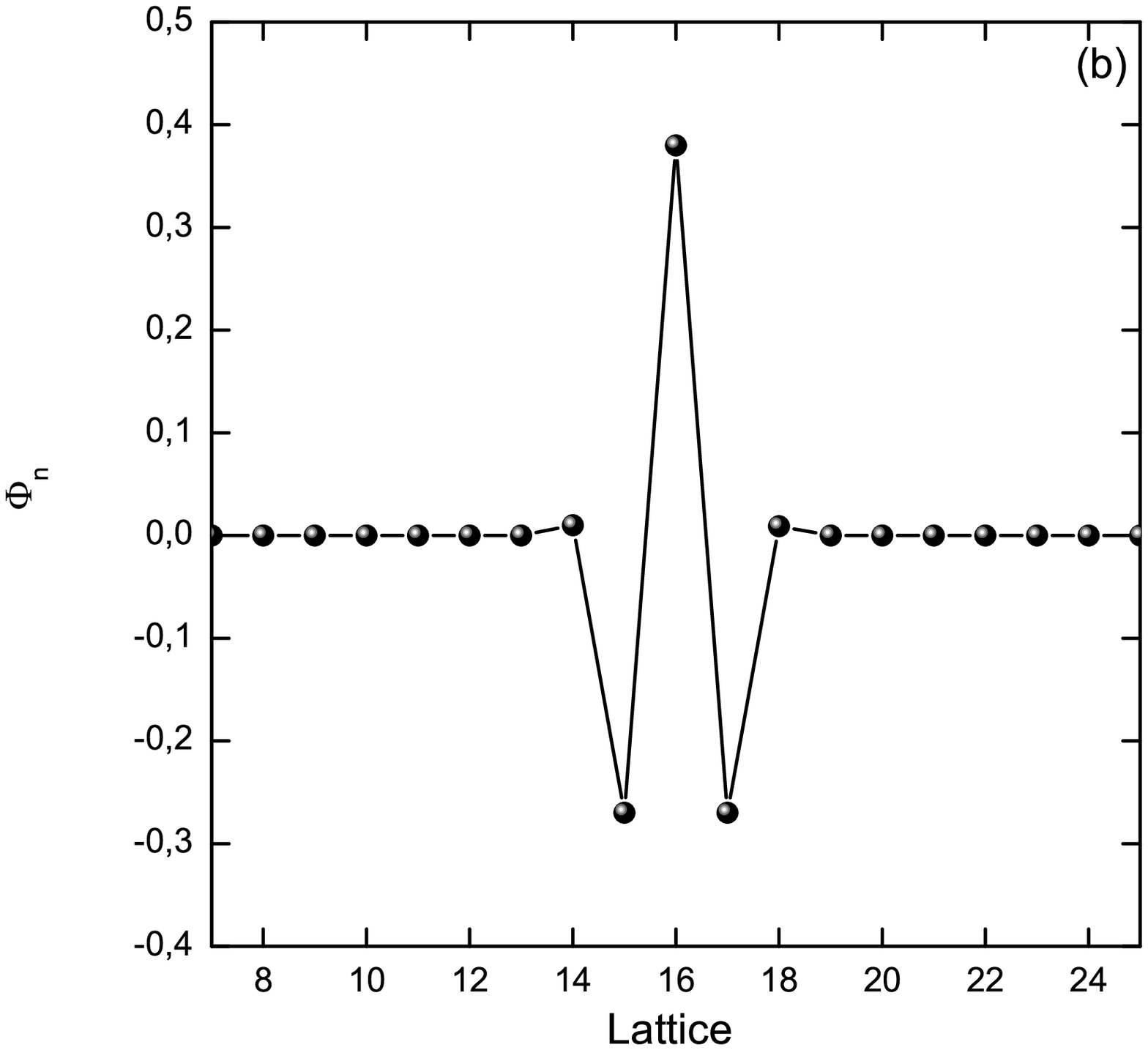}
  \caption{(a) Initial conditions for the 31 particle lattice, obtained from homoclinic intersections of the spatial
   map $\Phi_{n}$ with all velocities set to zero. Here $D=1$, representing a quasiperiodic breather (Case I in the text).
   (b) We now perturb the initial conditions, so that the amplitudes of the central particle and its 2 neighbors
   are slightly different than those in (a) (Case II in the text).}\vspace{0.5cm}
   \label{initials}
\end{figure*}
%=============================================
Let us take, for example, a Hamiltonian lattice of anharmonic
oscillators, which involves only quartic coupling terms and hence
presents strong localization phenomena due to the absence of phonons
\cite{Gor_Flach}. More specifically, we study here a system with on
site potential: $V(x)=\frac{1}{2} [1-\varepsilon \cos (\omega
_{d}t)]x^{2}-\frac{1}{4}x^{4}$, described by the Hamiltonian
\cite{Man_Bou}:\vspace{-0.25cm}
\begin{equation}
H(t)=\underset{n}{\sum }\{\frac{1}{2}\dot{x}_{n}^{2}+\frac{1}{2}
[1-\varepsilon \cos (\omega
_{d}t)]x_{n}^{2}-\frac{1}{4}x_{n}^{4}+\frac{K}{4}
(x_{n+1}-x_{n})^{4}\}, \label{ham}\vspace{-0.25cm}
\end{equation}
where $\varepsilon$, $\omega_d$  are the amplitude and frequency
of the driver respectively.
%=============================================
\begin{figure*}[h]
  \includegraphics[width=8.5cm]{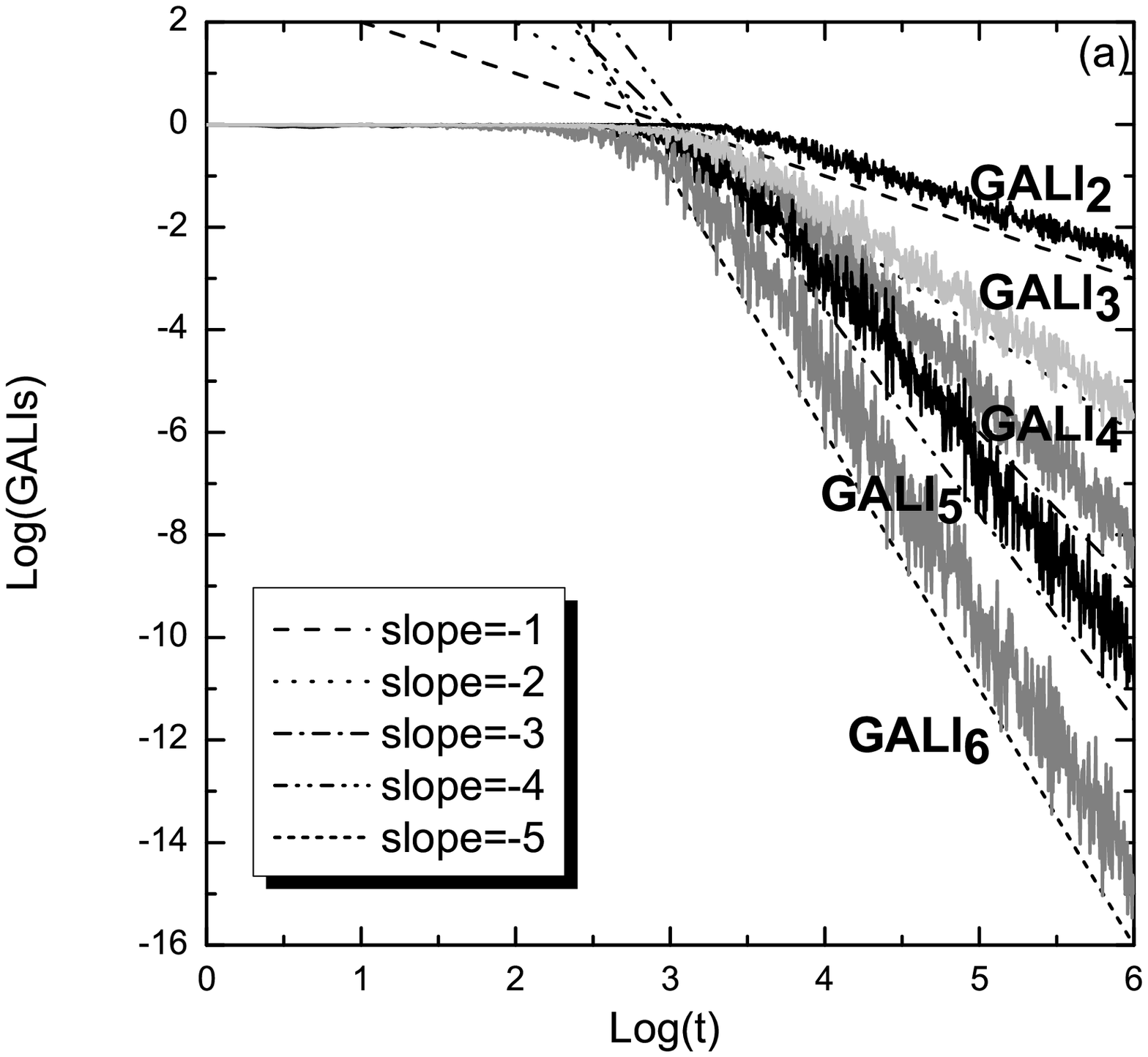}\hspace{-1.5cm}
  \includegraphics[width=8.5cm]{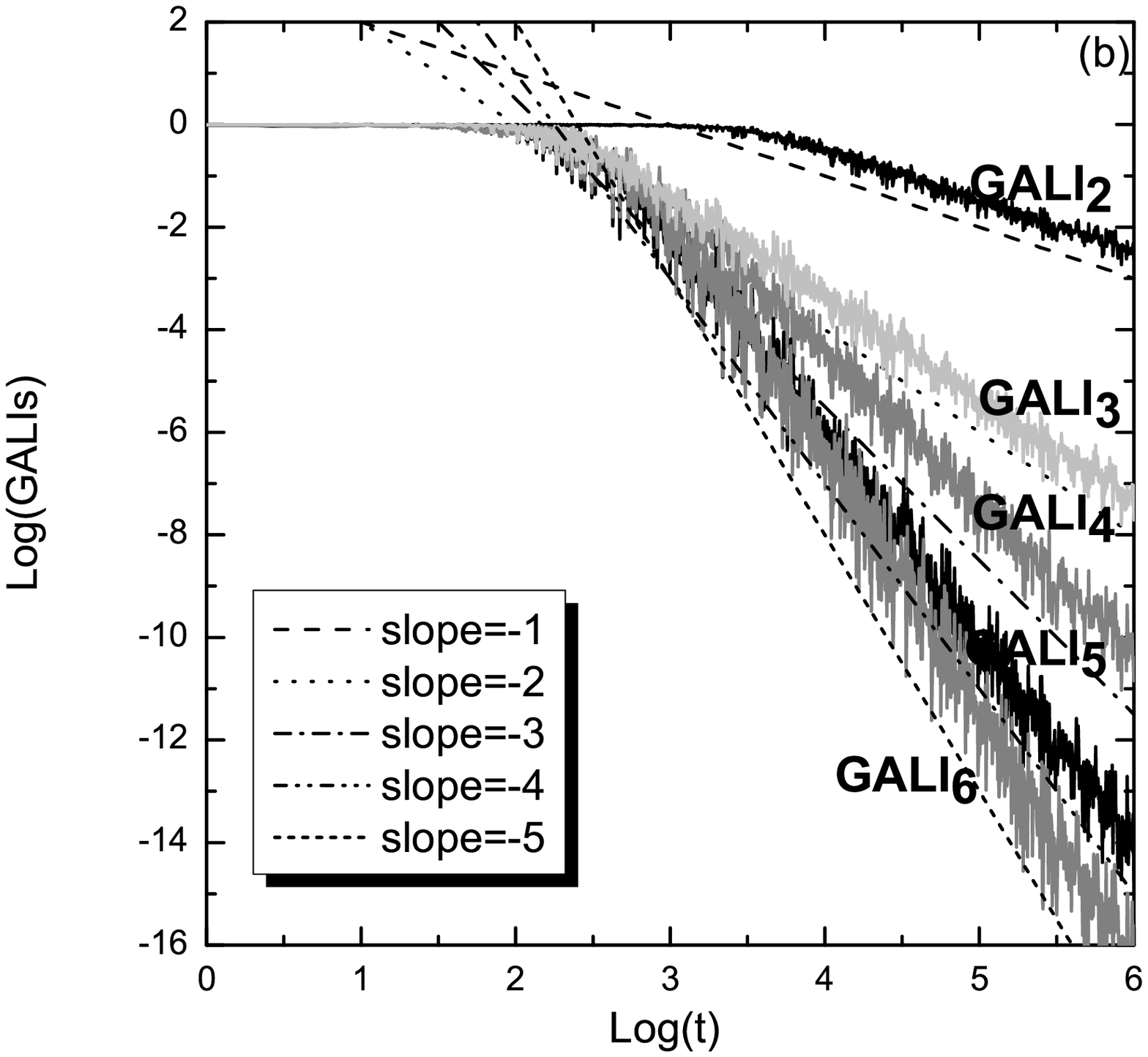}\vspace{-0.5cm}
  \includegraphics[width=8.5cm]{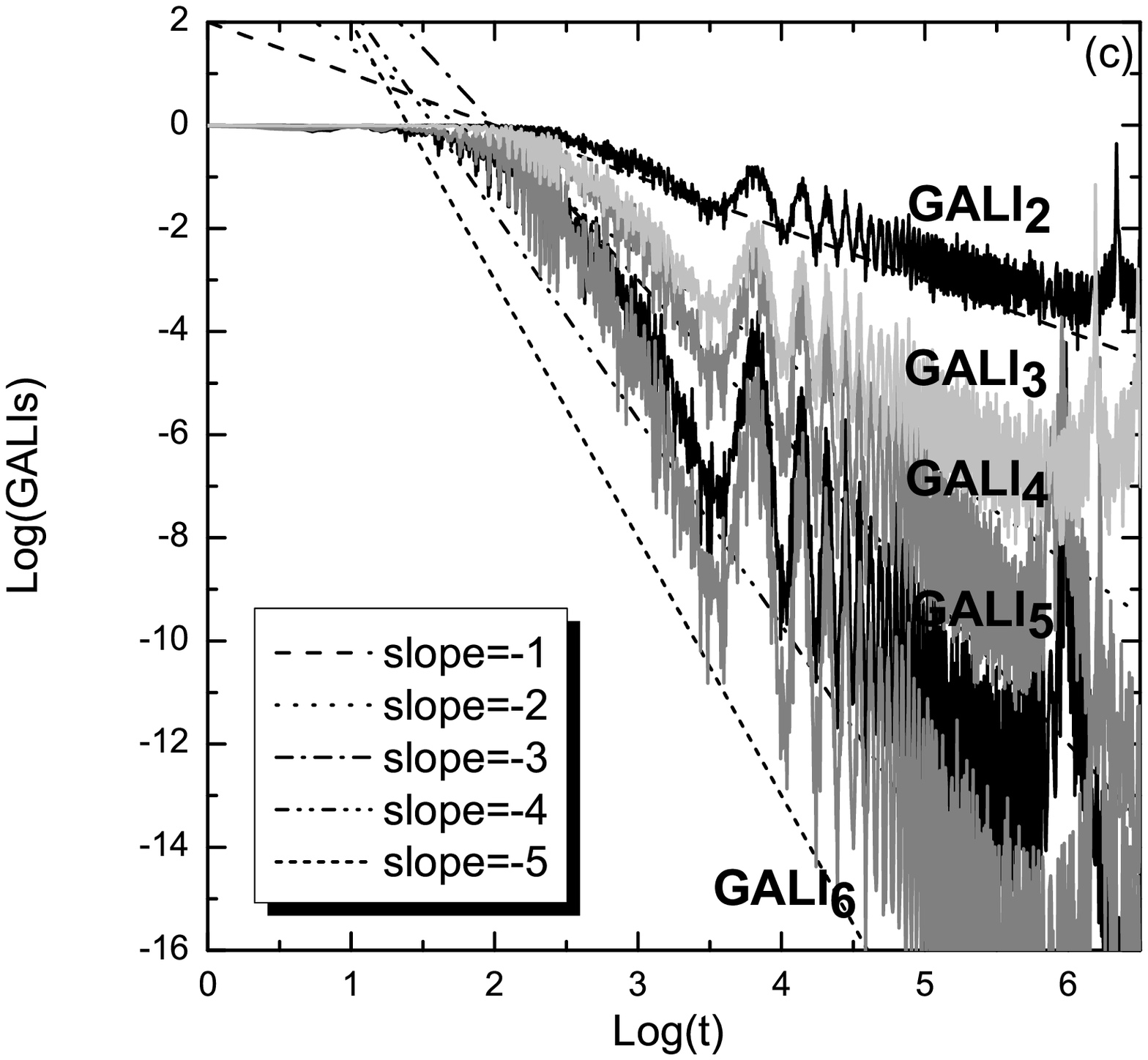}\hspace{-1.5cm}
  \includegraphics[width=8.5cm]{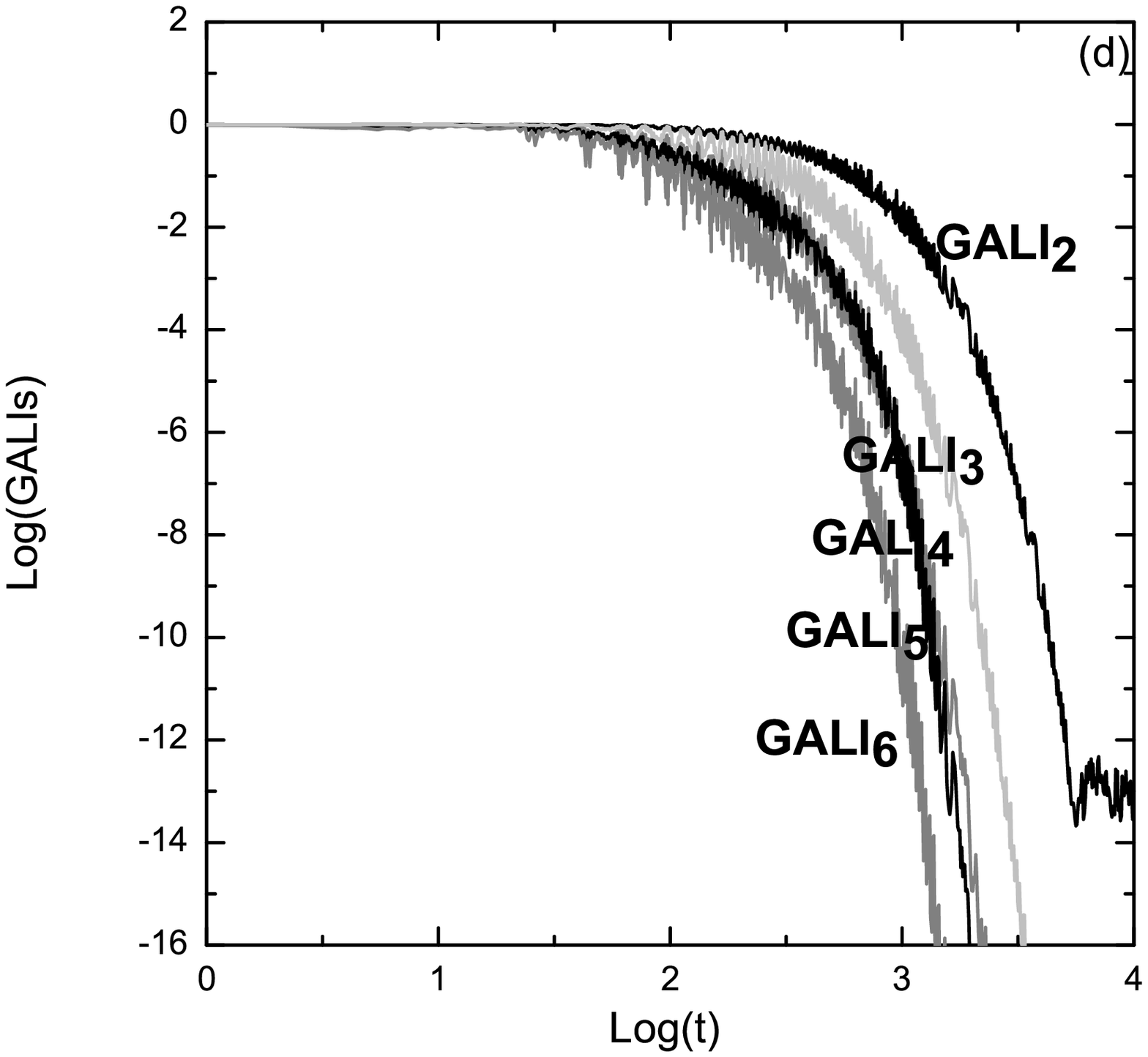}
  \caption{Time evolution of GALI$_{k}$ indices for: (a) $D=1.1$ and initial energy $H(0)=0.133$, see (\ref{ham}), (b) $D=1$ and $H(0)=0.097$,
  (c) $D=0.9$ and $H(0)=0.07$ and (d) for  $D=0.8$ and $H(0)=0.048$. Notice that in (a) and (b) the breather
  remains quasiperiodic for very long times, lying on a one--dimensional torus, which implies that only one main frequency is
  excited. In (d) the motion is evidently chaotic and delocalizes, drifting away from the breather.
  The slopes of GALI$_{k}$ evolution coincide with the laws presented in formulas (\ref{GALI:1}) and
  (\ref{GALI:2}).}
  \label{case:1}
\vspace{0.5cm}
\end{figure*}
%=============================================
The equations of motion of each oscillator clearly are:
\begin{equation}
\ddot{x}_{n}=K(x_{n+1}-x_{n})^{3}+K(x_{n-1}-x_{n})^{3}-[1-\varepsilon
\cos (\omega_{d}t)]x_{n}+x_{n}^{3}. \label{2}
\end{equation}
It can be shown that the solutions of equation (\ref{2}) can be
written as a product of a time--dependent function $G(t)$ and a
map $\Phi_{n}$, depending only on the oscillation site $n$:
\begin{equation}
x_{n}(t)=\Phi_{n}G(t), \label{3}
\end{equation}
$n=1,2,...,N$. Substituting (\ref{3}) in (\ref{2}), one finds that
the equations of motion can be exactly separated into a spatial
and a time--dependent part each of which is equal to an arbitrary
constant $C$. Thus, we arrive at a periodically driven Duffing
equation:
\begin{equation}
\ddot{G}+[1-\varepsilon \cos (\omega_{d}t)]G=-CG^{3}, \label{duff}
\end{equation}
satisfied by $G(t)$ and a discrete map:
\begin{equation}
C\Phi_{n}+K(\Phi_{n+1}-\Phi_{n})^{3}+K(\Phi_{n-1}-\Phi_{n})^{3}+\Phi
_{n}^{3}=0, \label{discr}
\end{equation}
which yields the spatial evolution of the system (\ref{ham}).
Following \cite{Man_Bou}, we consider a lattice of $31$ particles,
with $C=1,K=1,\varepsilon =0.7$ and $\omega_{d}=2.6355$. The initial
conditions of the driven Duffing equation are chosen on an exact
exact periodic orbit with $G(0)=D$, $\dot{G}(0)=0$ and frequency
$\omega_b=\omega_d/2$, if $D=D_b=\pm 1.2043$. Furthermore, if the
initial conditions lie on the \textit{homoclinic intersections} of
the invariant manifolds of the saddle point at the origin of the
map, the solution becomes an exact breather with
$x_{n}(0)=\Phi_{n}D_b$ (Fig. \ref{initials}a), which is
\textit{linearly stable} \cite{Man_Bou} .

Our first goal, therefore, is to examine the motion near this
breather by perturbing the initial conditions in its neighborhood
and applying the GALI indices to characterize the resulting orbits.
In this study, we have made two kinds of perturbations: In the first
case we change only the factor $D$, while in the second both the
factor $D$ and the spatial initial conditions of the map are varied.
%=============================================
\begin{figure*}[t]
  \includegraphics[width=8.5cm]{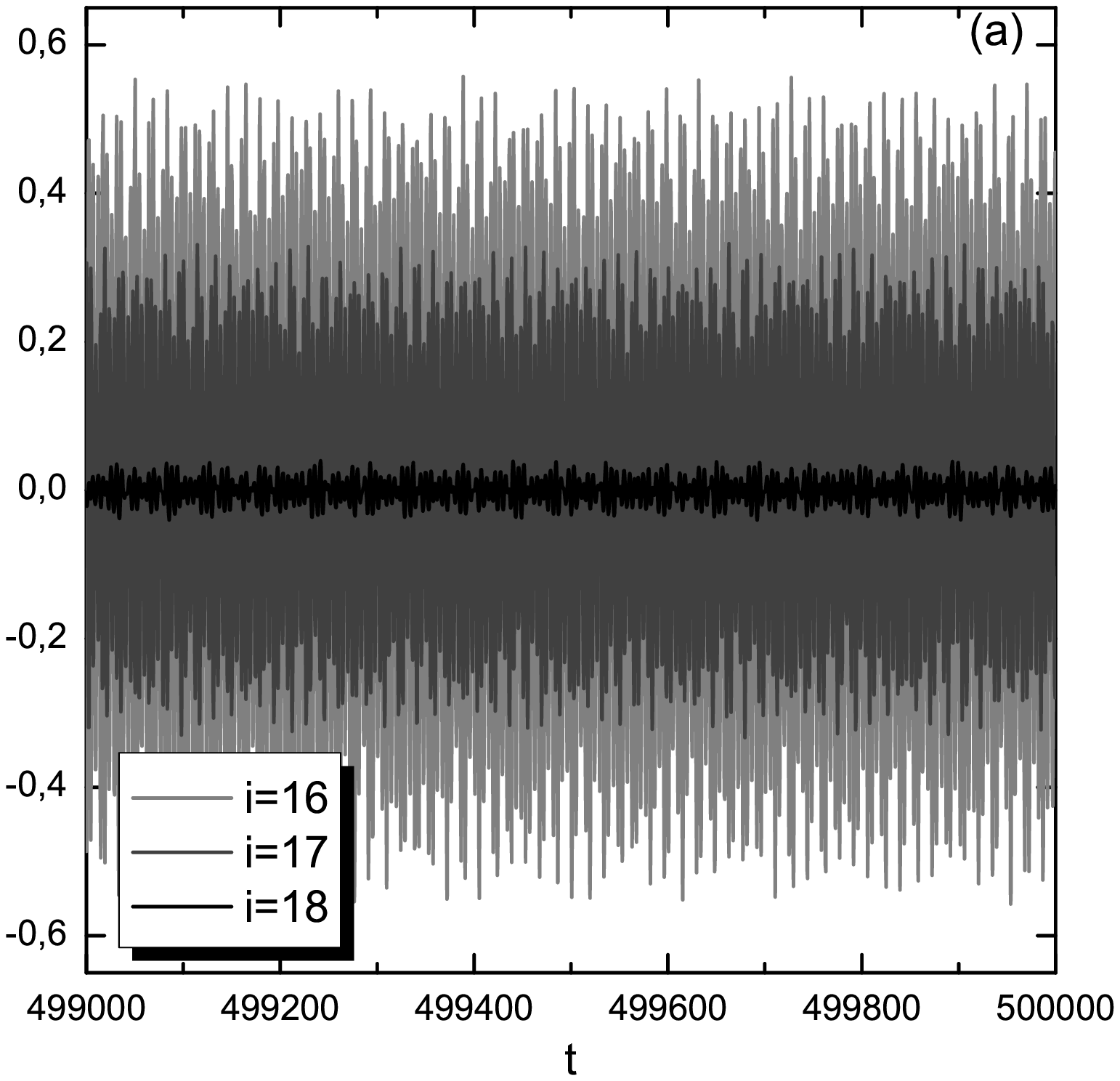}\hspace{-1.5cm}
  \includegraphics[width=8.5cm]{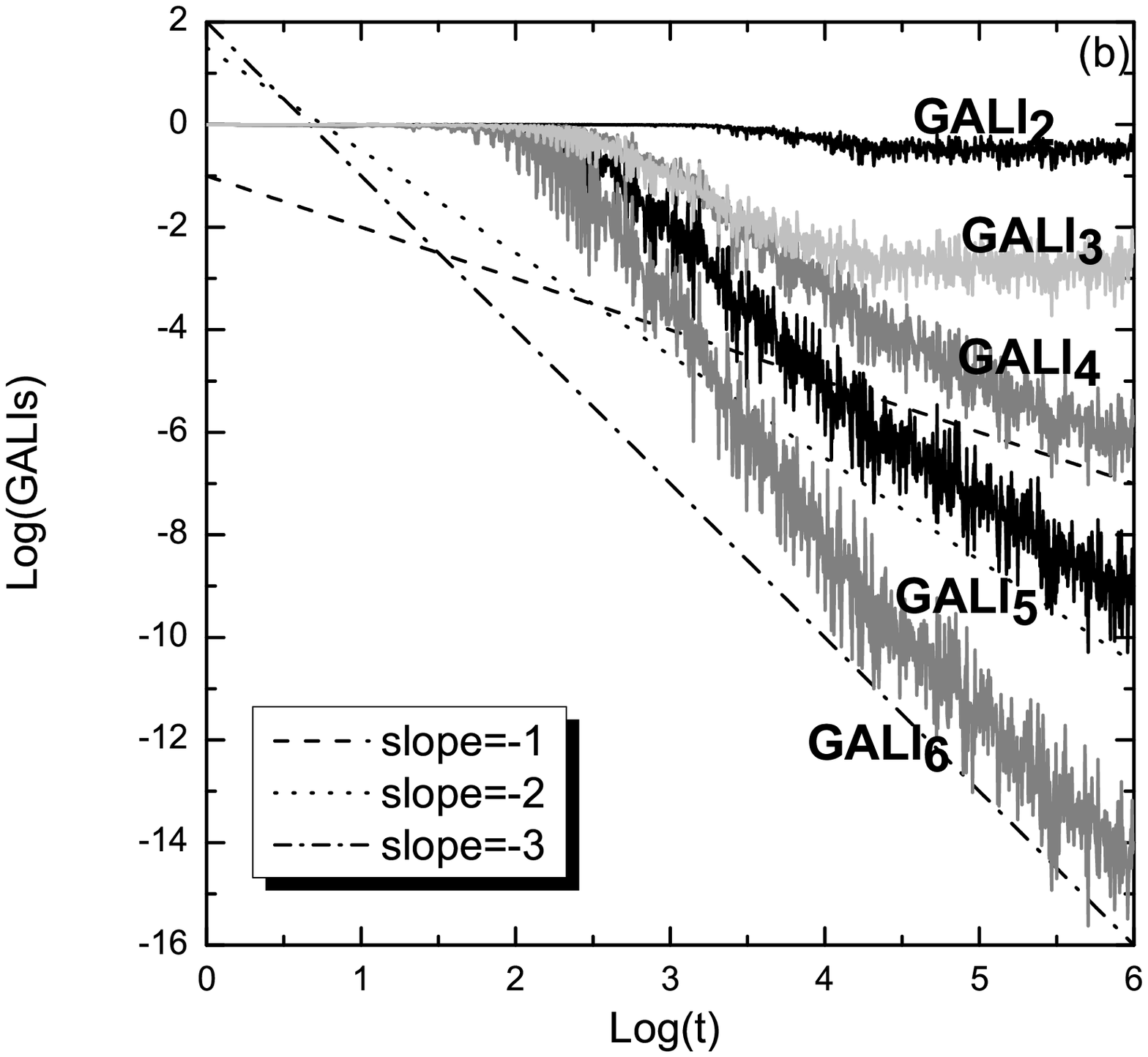}\vspace{-0.5cm}
  \includegraphics[width=8.5cm]{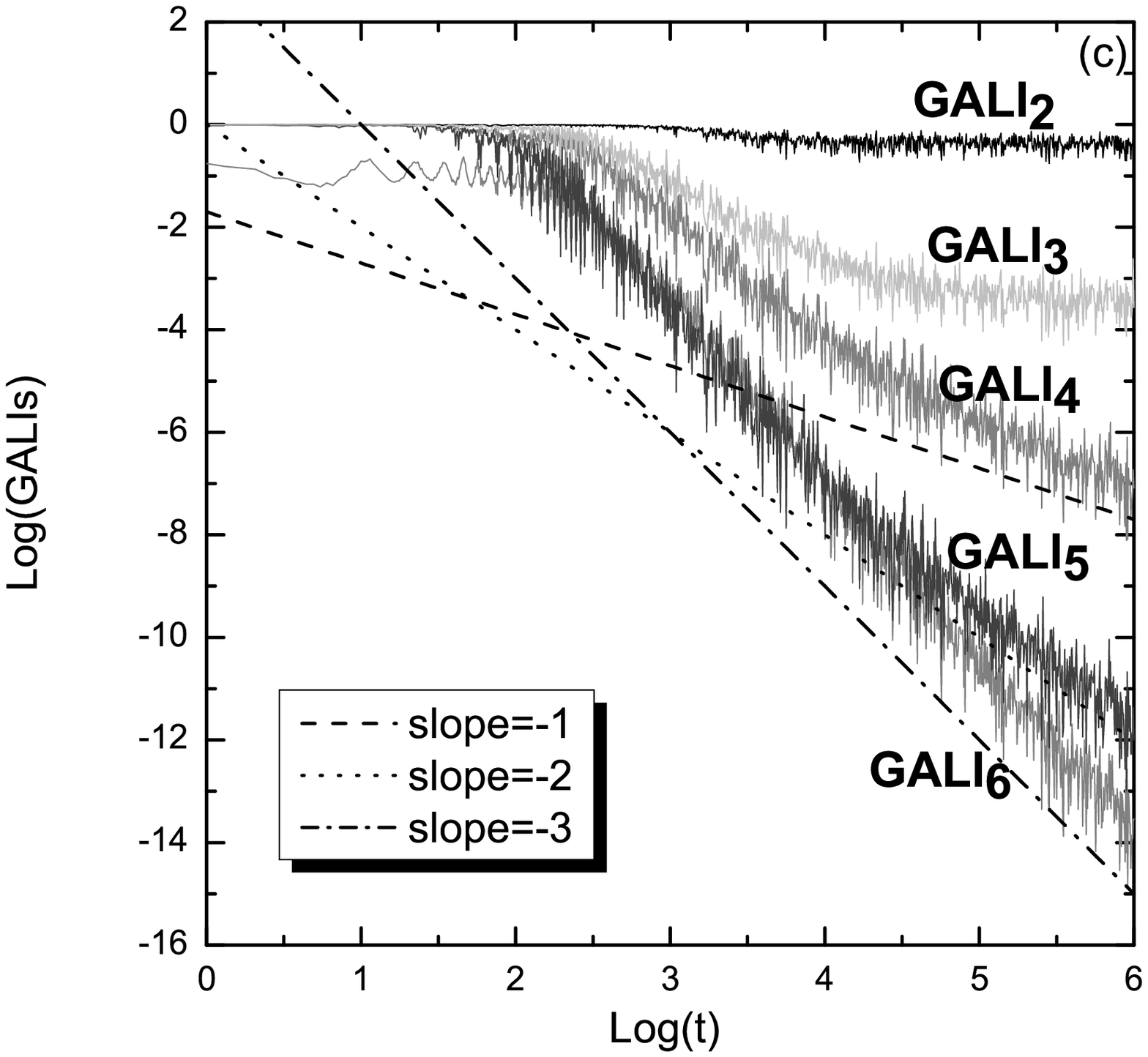}\hspace{-1.5cm}
  \includegraphics[width=8.5cm]{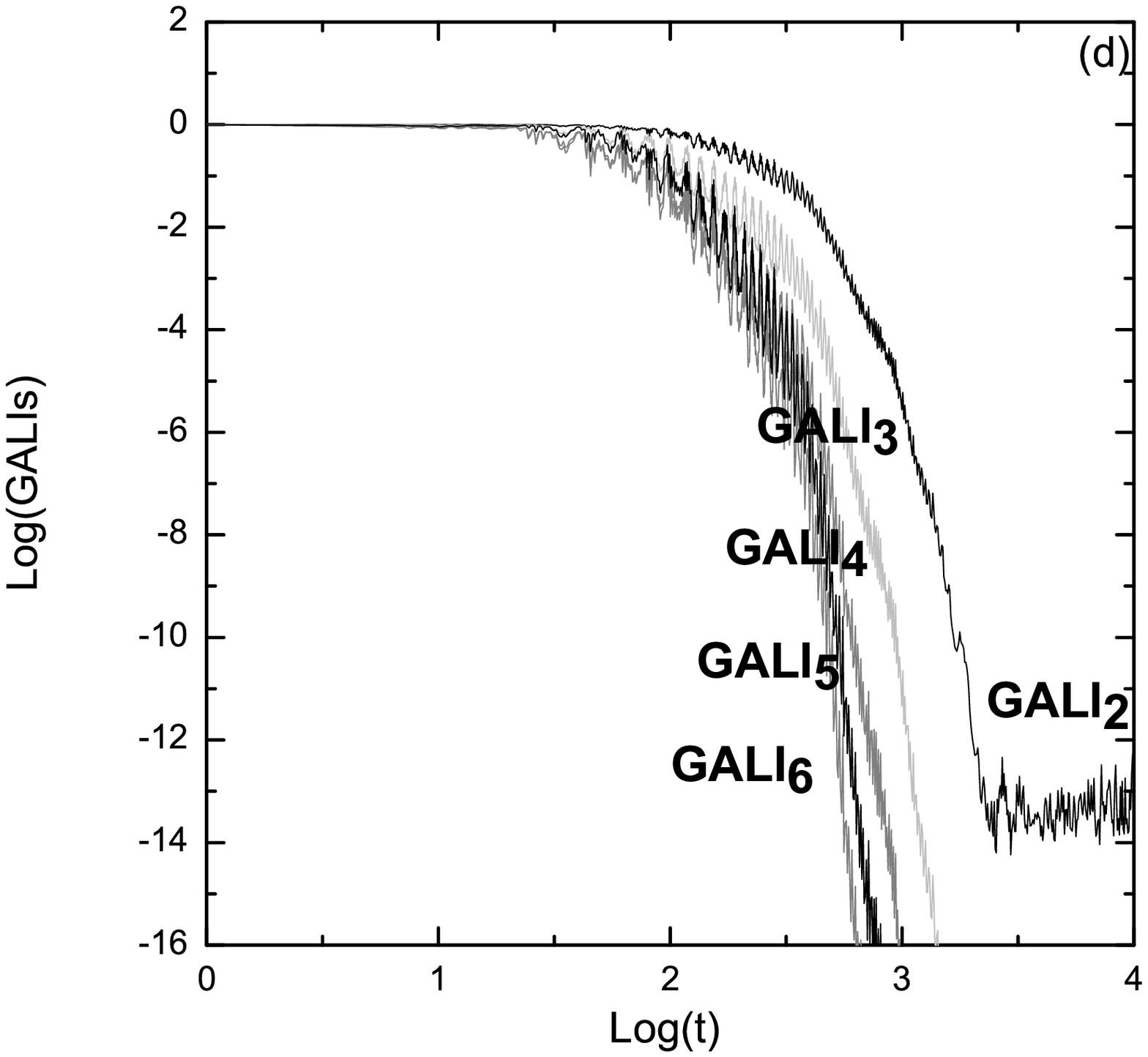}
  \caption{(a) Oscillations of particles 16, 17, 18 of the lattice show the localization and quasiperiodic properties
   of the motion near the discrete breather. This is verified by the time evolution of GALI$_{k}$ indices in (b) for $D=0.9$
  and $H(0)=0.09$ and in (c) for $D=0.8$ and $H(0)=0.062$, showing the orbits lie on a 3--dimensional torus. However, in (d)
  we take $D=0.7$ and $H_{0}=0.042$, perturbing the initial conditions as shown in Fig. \ref{initials}b.
  Here, the slopes of the GALI$_{k}$ indices clearly demonstrate that the motion chaotically diffuses away.}\vspace{0.5cm}
\label{case:2}
\end{figure*}
%=============================================

\textbf{Case I}:\ \vspace{-0.5cm}

Let us start by perturbing the initial conditions $G(0)=D$
($\dot{G}(0)=0$) away from $D=D_{b}=1.2043,$, while the initial
conditions of $\Phi_{n}$ are fixed (Fig. \ref{initials}a). In our
example, we use a $31$ particle Hamiltonian, with $C=1$,
$\varepsilon =0.7$ and $\omega_{d}=2.6355$.

Decreasing the value of $D<D_{b}$, we observe that the oscillations
for $D=1.1$ are quasiperiodic and remain so for long integration
times, while the particles oscillate with frequencies close to that
of the exact breather, as shown in Fig. \ref{case:1}. This is in
agreement with the fact that even GALI$_{2}$ (representing the area
of the parallelogram formed by the two unit deviation vectors) tends
to zero following the power law $t^{-1}$. When $D$ becomes 0.9, the
GALI$_{k}$ indices begin to fluctuate, indicating that the orbit is
near the edge of the corresponding regular region in phase space
(Fig. \ref{case:1}c). Moreover, near $t=10^{6}$, GALI$_{2}$ begins
to tend to a constant value, suggesting that the motion has become
quasiperiodic with 3 frequencies. Finally, as $D$ reaches the value
0.8, quasiperiodic recurrences break down and the motion becomes
chaotic.
%=============================================
\begin{figure}[t]
  \includegraphics[width=8.5cm]{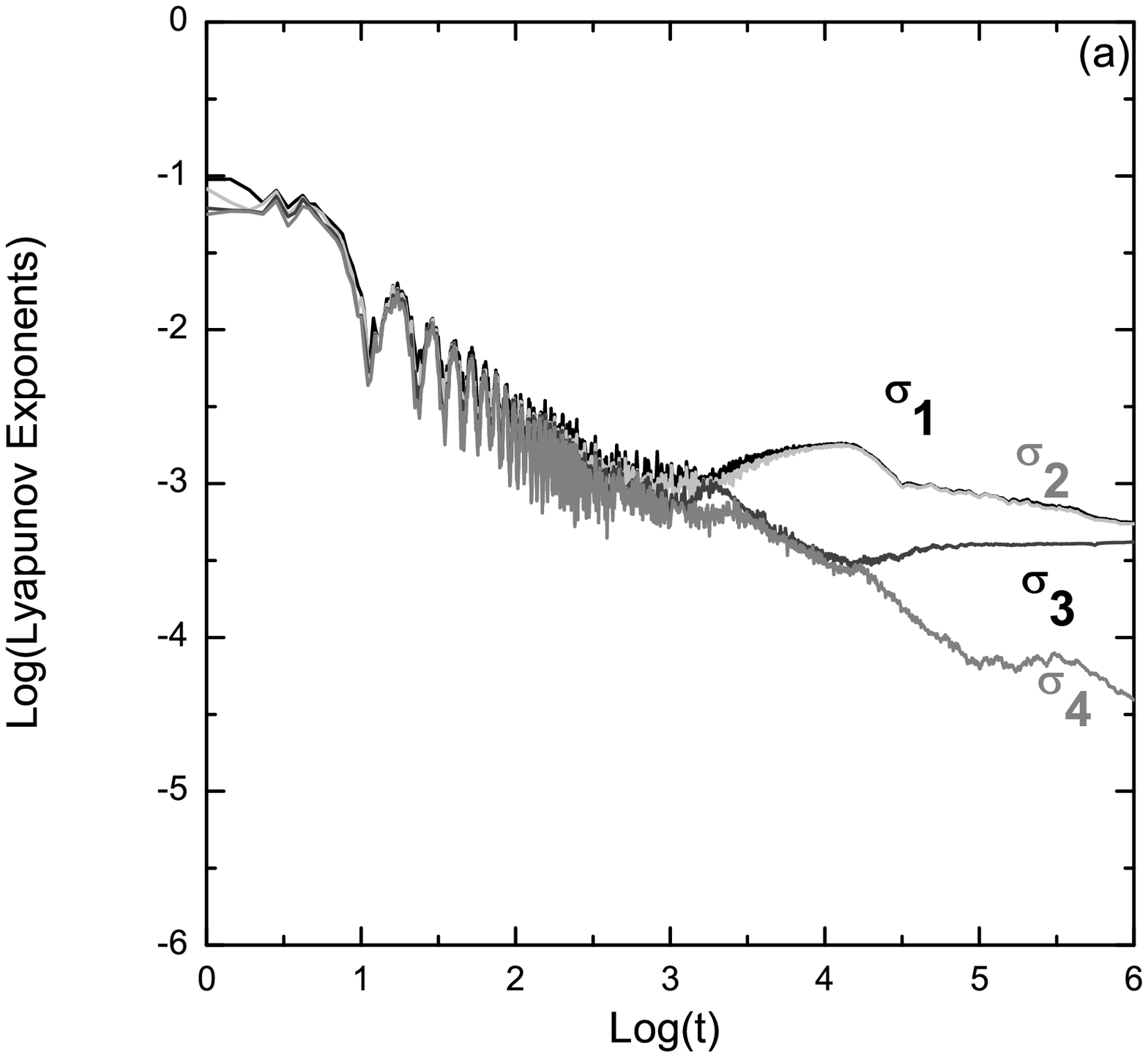} \hspace{-1.5cm}
  \includegraphics[width=8.5cm]{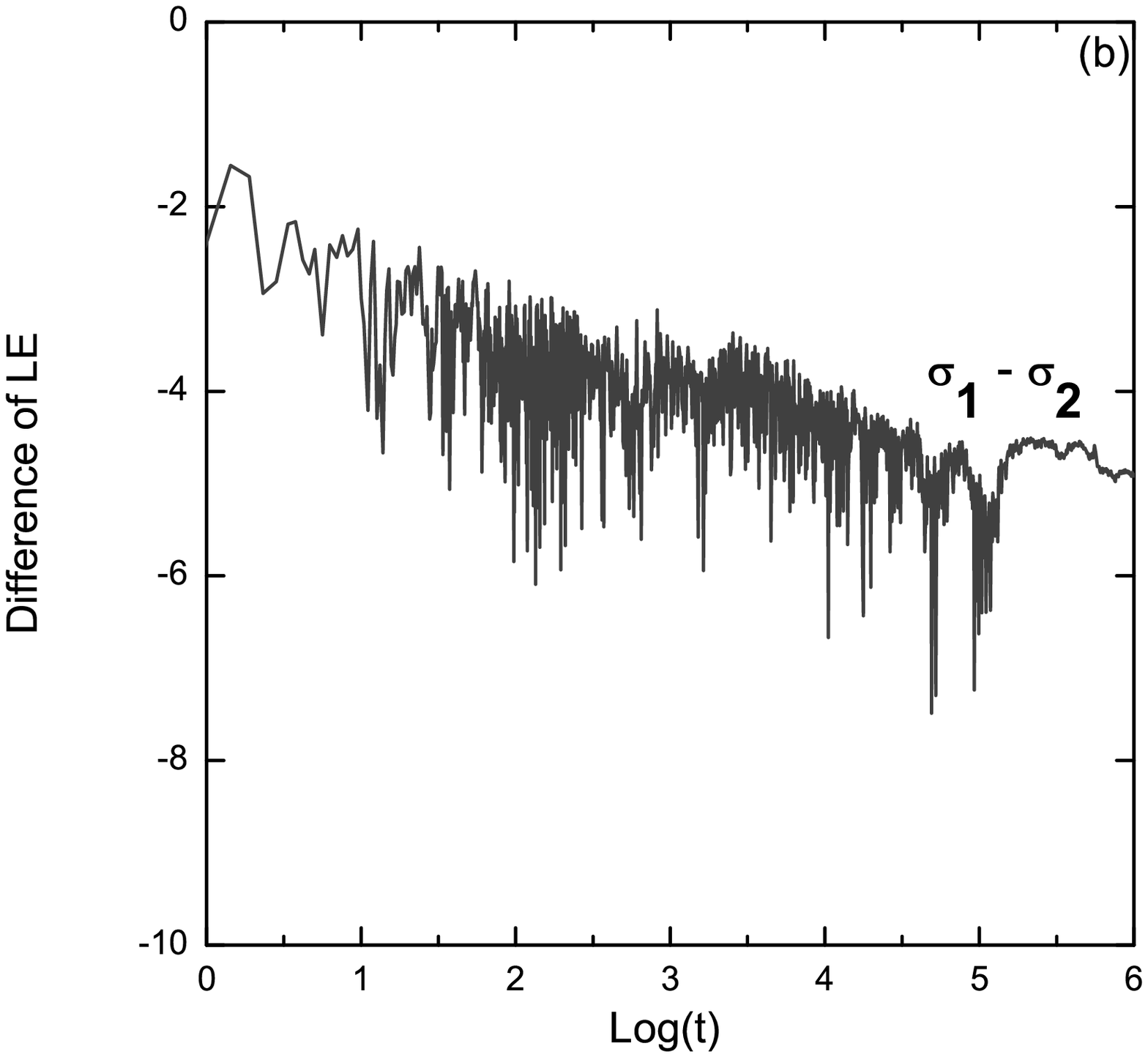}\vspace{-0.5cm}
  \includegraphics[width=8.5cm]{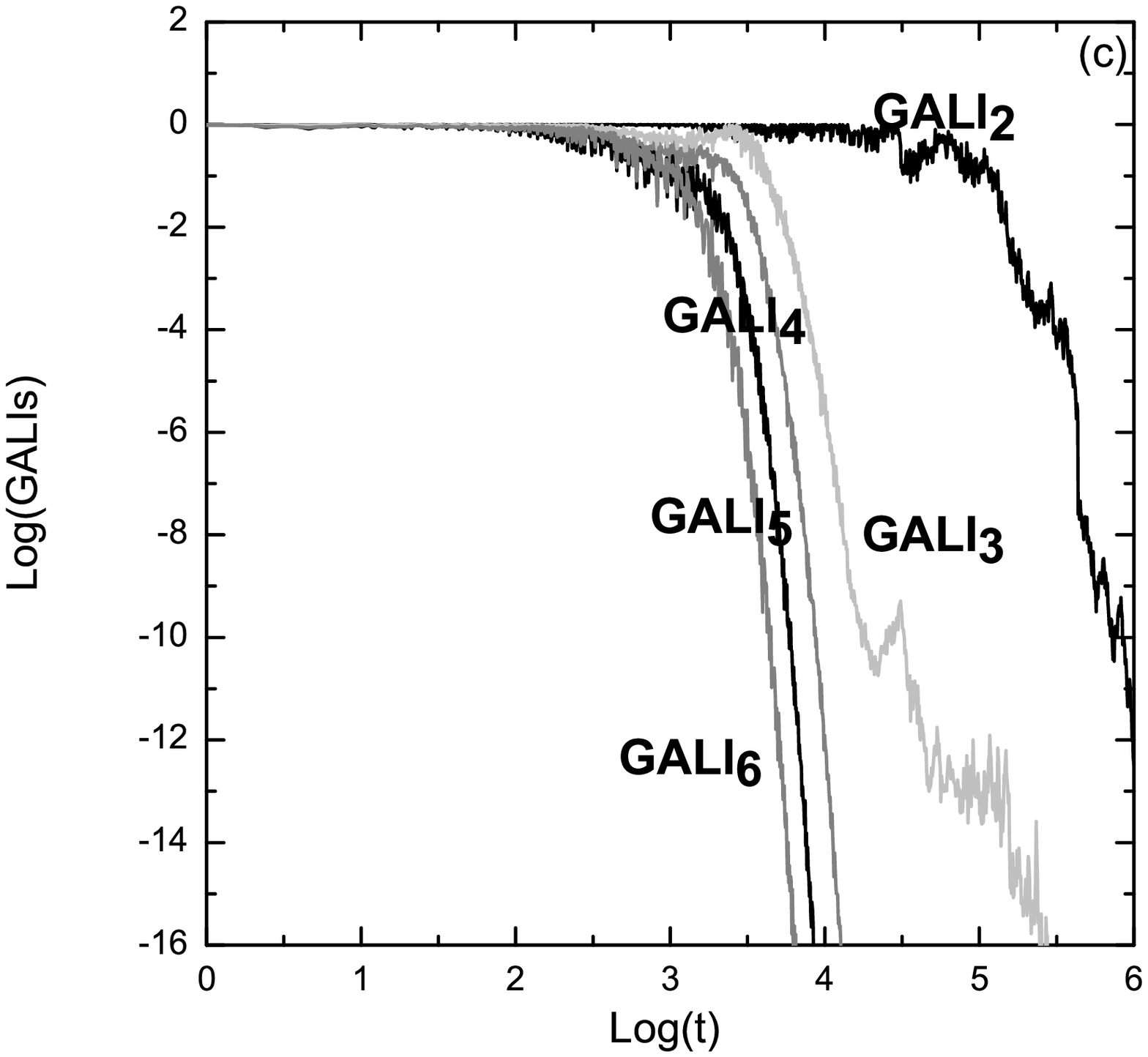} \hspace{-1.5cm}
  \includegraphics[width=8.5cm]{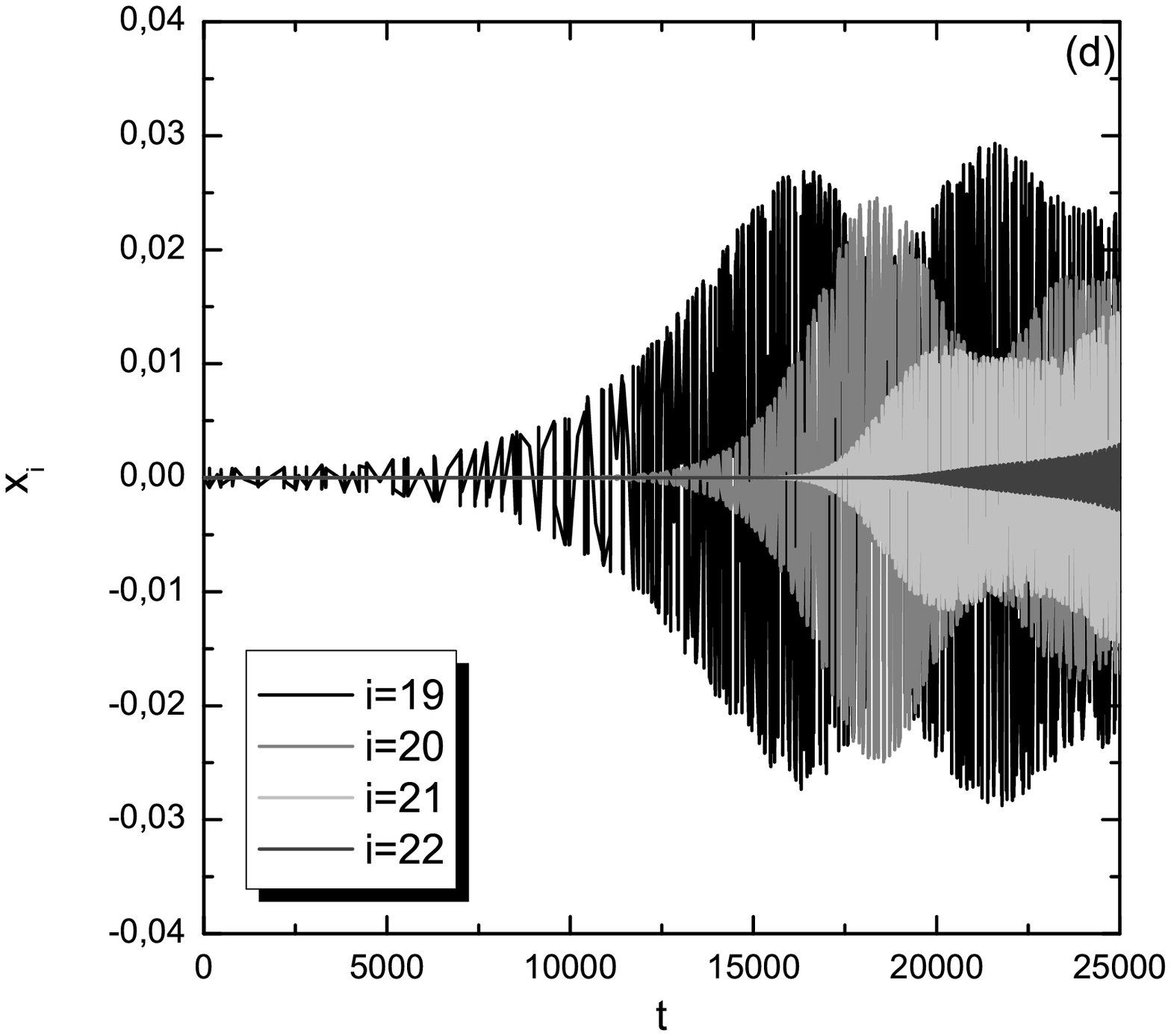}
  \caption{When we choose initial conditions as shown in Fig. \ref{initials}b (Case II), we find that with $D=1.2$
   recurrences break down, while (a) the four largest Lyapunov exponents appear to be decreasing to zero. Note
   in (b) that the difference between the two largest Lyapunov exponents, $\sigma_{1}-\sigma_{2}$, is nonzero and tends
   to a constant, explaining why in (c) we observe exponential behavior of GALI$_{k}$. (d) This agrees with the fact that
   the particles away from  the center of the lattice start to oscillate,
   indicating delocalization and the breakdown of recurrency.}\vspace{0.5cm}
   \label{break}
\end{figure}
%=============================================

\textbf{Case II}:\ \vspace{-0.5cm}

We now perturb slightly the initial conditions of the discrete map
$\Phi_{n}$ (as indicated in Fig. \ref{initials}b) and observe that
the motion now occurs on a 3D torus, indicating the appearance of 3
frequencies in the system. This quasiperiodic breather behavior lies
on a 3D torus, in phase space, as $D$ varies from 0.75 to 1 (Fig.
\ref{case:2}). In fact, for $1<D<1.2$, the chaotic character of the
orbit is revealed by the exponential decay of all GALI$_{k}$
indices.

For $D=1.2$, in fact, recurrences break down and delocalization sets
in, as particles away from the center of the lattice gain energy and
begin to oscillate (see Fig. \ref{break}d). The exponential
convergence of GALI indices to zero indicates the chaotic behavior
of the motion. Fig. \ref{break}a shows that there are only three
positive Lyapunov exponents, indicating three instability directions
about the orbit in phase space. The exponents $\sigma_{1}$ and
$\sigma_{2}$ are almost identical, having a difference of the order
$10^{-5}$ (Fig.\ref{break}b). As a consequence, GALI$_{2}\propto
e^{-(\sigma_{1}-\sigma_{2})t}$ begins to converge exponentially when
$t$ becomes greater than $10^{-5}$, while the other GALIs have
reached the value $10^{-16}$ exponentially, long before $10^{4}$
(see Fig. \ref{break}c).

%=============================================
\vspace{-0.5cm}
\section{Application to the dynamics of $N$ coupled standard maps}
\vspace{-0.75cm} \label{N_sm}

The standard map, also referred to as the Chirikov--Taylor map
\cite{mac}, is an area preserving map, which appears in many
physical problems. It is defined by the equations:
\vspace{-0.75cm}
\begin{eqnarray}
  x_{n+1} &=& x_n + y_{n+1}, \qquad  \\ \nonumber
  y_{n+1} &=& y_n + K \sin (x_n),
\end{eqnarray}
where both variables are taken modulo one in the unit square. This
map describes the motion of a simple mechanical system called a
kicked rotator. It may be thought of as representing a pendulum
rotating on a horizontal frictionless plane around a fixed axis and
being periodically kicked by a nonlinear force at the other end, at
unit time intervals. The variables $x_n$ and $y_n$ determine,
respectively, the angular position of the pendulum and its angular
momentum after the $n^{th}$ kick. The constant $K$ measures the
intensity of the nonlinear ``kicks".

The standard map describes many other systems occurring in the
fields of mechanics, accelerator physics, plasma physics, and solid
state physics. However, it is also interesting from a fundamental
point of view, since it is a very simple model of a Hamiltonian
system of 2 degrees of freedom that displays order chaos.

\begin{figure*}[t]
\begin{center}
  \includegraphics[width=8.5cm]{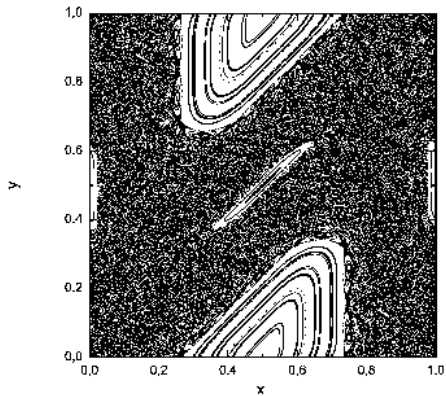}
  \caption{Poincar\'{e} surface of section of a single standard map for
  $K=2$. Note the presence of a stable fixed point at $(x,y)=(0.5,0.0)$ (and its modulo equivalent at $(x,y)=(0.5,1.0)$),
  surrounded by a large ``island" of quasiperiodic orbits. The smaller islands embedded in the large chaotic
  ``sea" correspond to a stable periodic orbit of period 2.}\vspace{0.5cm}
  \label{pss}
\end{center}
\end{figure*}

For $K=0$, the map is linear and only periodic and quasiperiodic
orbits are allowed (depending on the value of $y_0$), which fill
horizontal straight lines in the $x_n, y_n$ phase plane of the map.
However, when $K\neq 0$, the periodic lines break into pairs of
isolated points half of which are stable and half unstable
alternatingly. The stable periodic points possess families of
quasiperiodic orbits forming closed curves around them, while the
unstable ones are saddles with thin chaotic layers close to their
invariant manifolds. Moreover, as $K>0$ grows, the size of the
chaotic layers increase and occupy larger and larger regions of the
available phase plane \cite{mac,Man:4}, see Fig. \ref{pss} above.

Let us consider a system of $N$ coupled standard maps described by
the following equations:\vspace{-0.75cm}
\begin{eqnarray}\label{eq:CSM}
  x_{n+1}^{j} &=& x_{n}^{j} + y_{n+1}^{j}, \\
  y_{n+1}^{j} &=& y_{n}^{j}+ \frac{K_j}{2\pi} \sin(2\pi x_{n}^{j})- \frac{\beta}{2\pi}\{ \sin[2\pi(x_{n}^{j+1}-x_{n}^{j})]+ \sin[2\pi(x_{n}^{j-1}-x_{n}^{j})]\} \nonumber
\end{eqnarray}
with $j=1,...,N$, fixed boundary conditions $x_0=x_{N+1}=0$ and
$\beta$ the coupling parameter between neighboring maps. In what
follows, we shall examine the localization properties of this system
comparing the dynamics with what is observed when one studies a
one--dimensional Hamiltonian system like the FPU lattice.
%=============================================
%=============================================

\textbf{A. Initial conditions localized in real space}
\vspace{-0.5cm}

Let us begin with the question of the existence of stable discrete
breathers in this model, when the coupling parameter $\beta$ is
small. Note that our equations have an exact stable fixed point,
when all particles are located at $(x,y)$ = (0.5,0.0) (see Fig.
\ref{pss}). Taking $N=20$ coupled standard maps, we may, therefore,
look for localized oscillations taking as initial condition (R1):
$(x_j,y_j)$ = (0.5,0.0), $\forall$ $j \neq 11$, perturbing only one
particle at $(x_{11},y_{11})=(0.65,0.0)$ and fixing the parameters
$\beta=0.001$, setting $K_j=2$, with $j=1,...,20$. If a stable
discrete breather exists, the orbits are expected to oscillate about
these initial conditions quasiperiodically for very long times.
%=============================================
\begin{figure*}[t]
  \includegraphics[width=8.5cm]{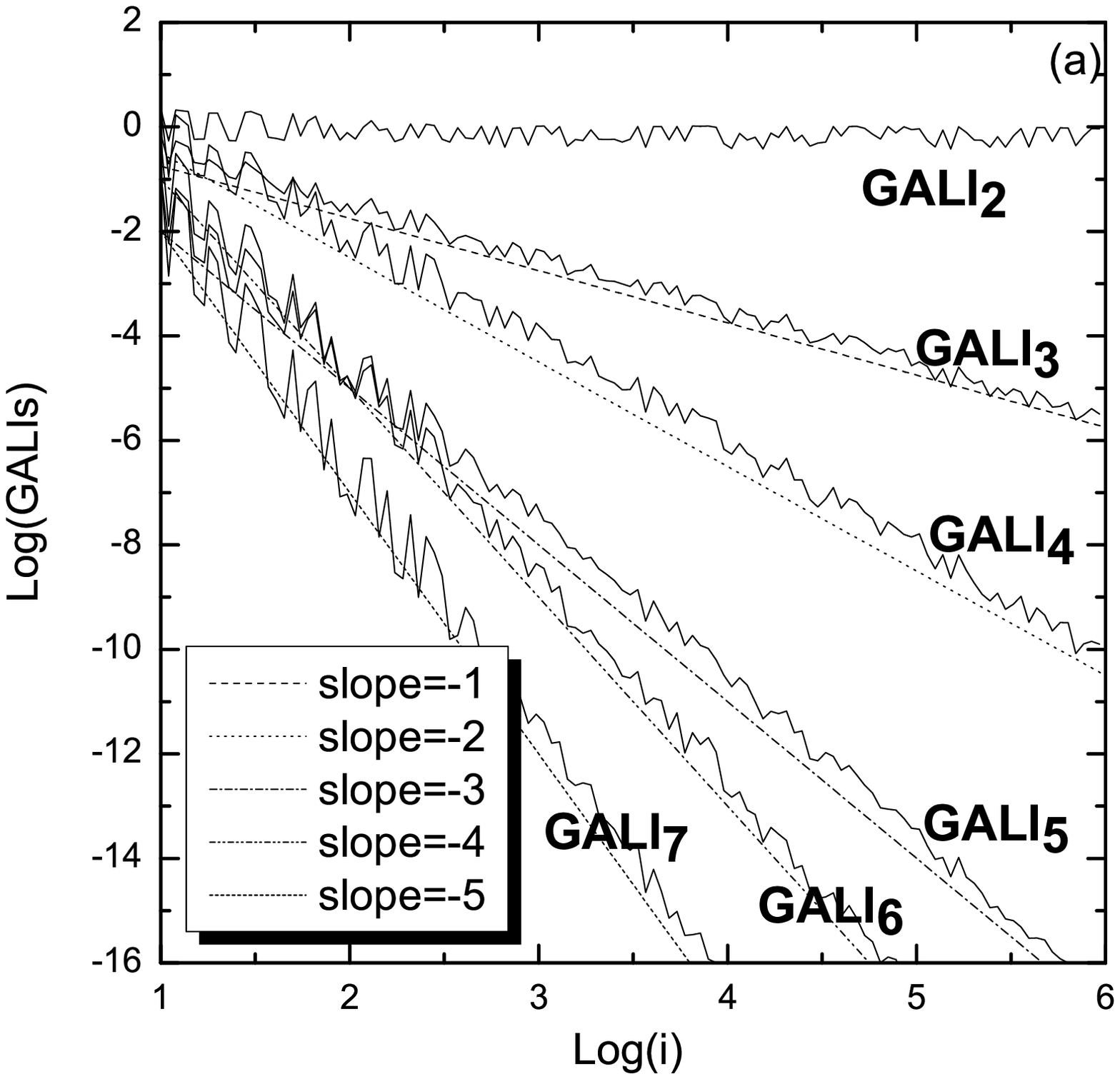}\hspace{-1.5cm}
  \includegraphics[width=8.5cm]{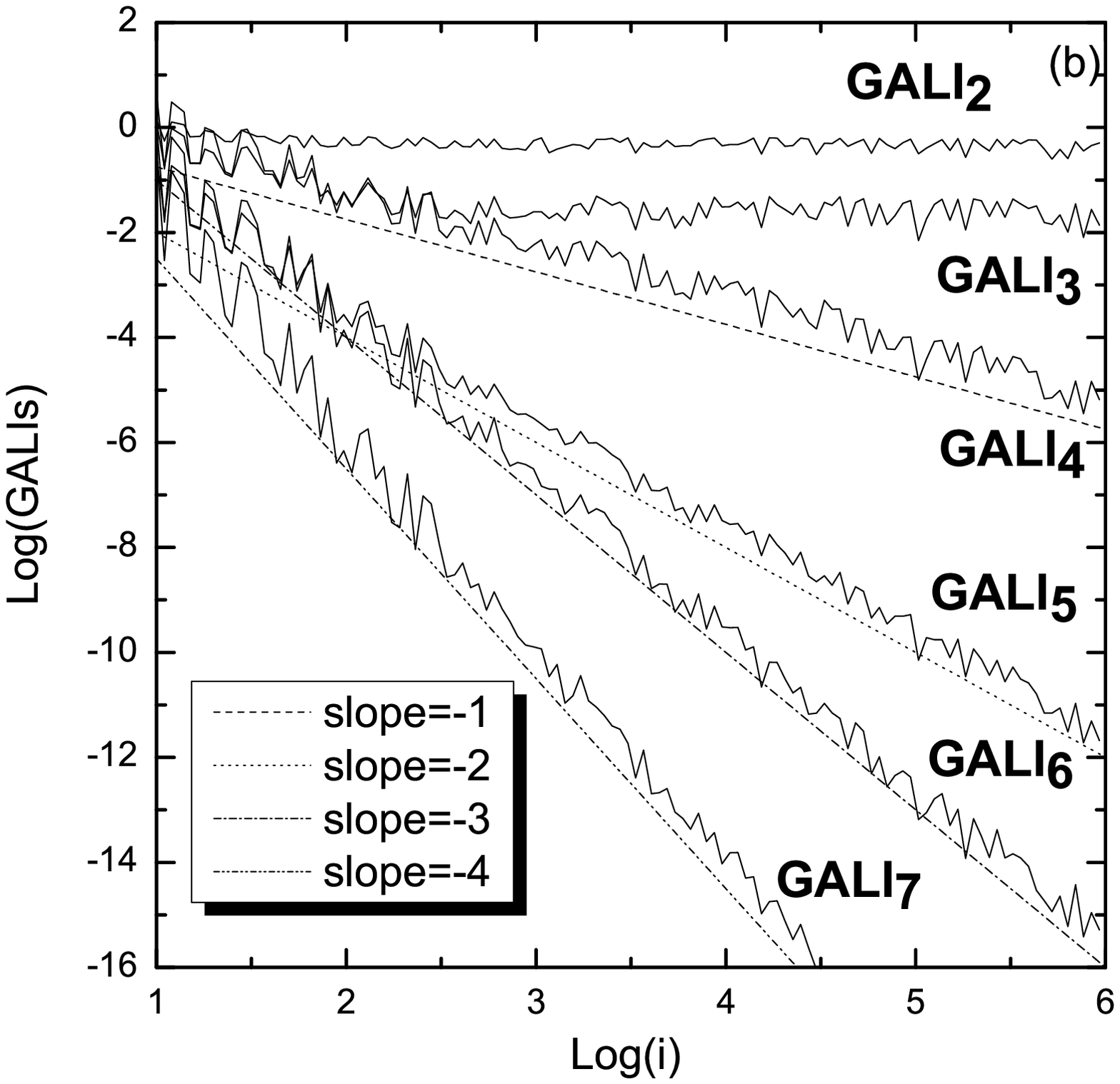}
  \begin{center}
 \hspace{1.25cm} \includegraphics[width=8.5cm]{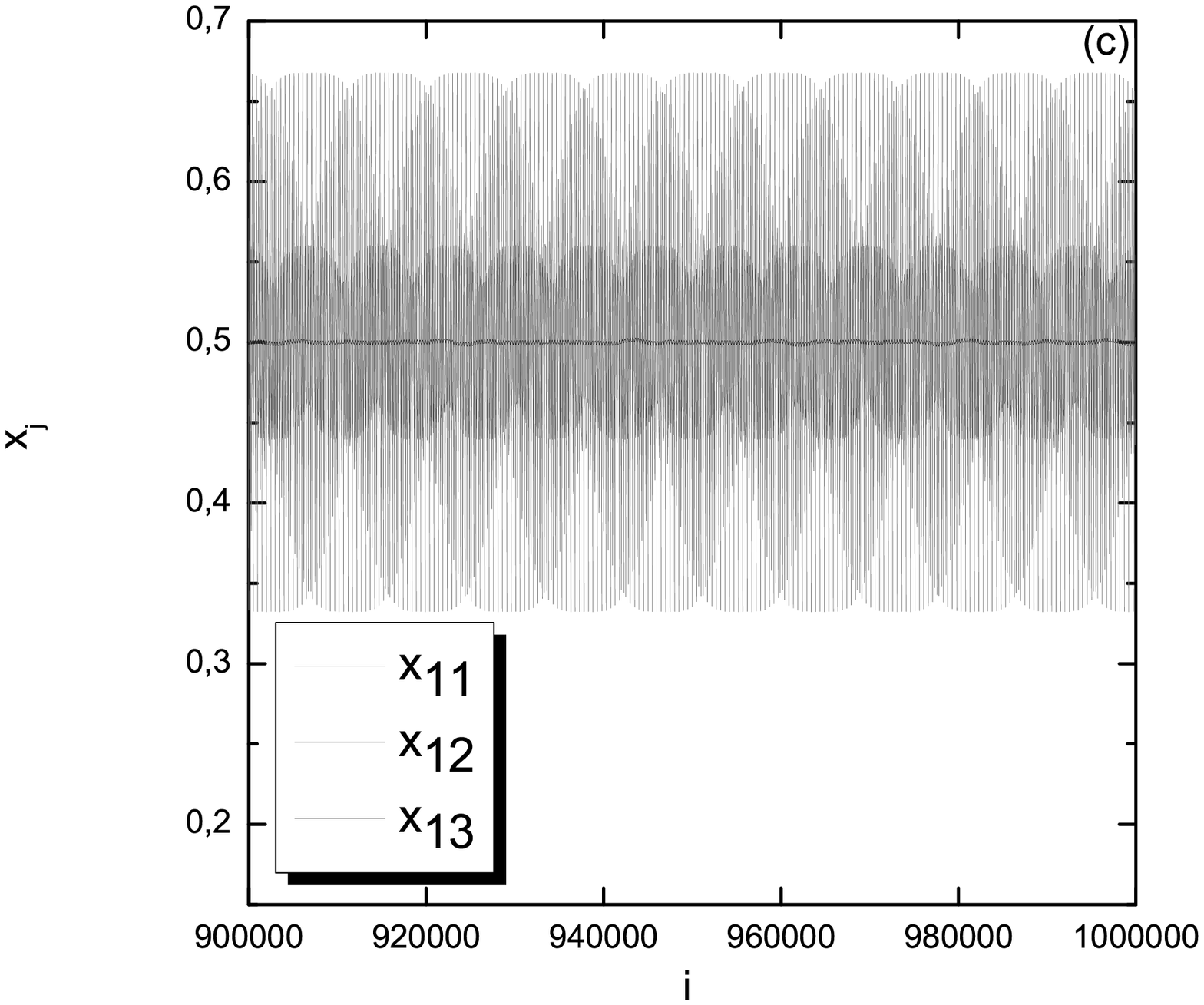}
  \end{center}
  \caption{(a) GALI$_k$, $k=2,...,7$, for for $N=20$ -- coupled standard maps with $K_j=2$, $\beta=0.001$ and initial
   conditions R1 (see text). GALI$_2$ fluctuates around a non--zero value implying a regular motion that lies on a 2D
   torus, while the slopes of the algebraically decaying indices are in agreement with (\ref{GALI:2}).(b) GALI$_k$ for
   the initial conditions R2 (see text) imply that the motion lies on a 3D torus, since GALI$_{2,3}$ tend to a non--zero
   number. This quasiperiodic localized dynamics of R2 is clearly seen in (c), where we plot the $x_n$ oscillations
   of the $11^{th}$  (light gray color), $12^{th}$ (gray color) and $13^{th}$ (black color) map. From the last $10^{5}$
   iterations shown in this figure, it is evident that the main part of the energy is confined only in the
   ``middle" 3 maps and is not shared by any of the other degrees of freedom of the system.}\vspace{0.5cm}
\label{1_2D_tor_real}
\end{figure*}

%=============================================
Indeed, when we evaluate the GALI$_k$ indices for $n=10^{6}$
iterations in Fig. \ref{1_2D_tor_real}a and display their evolution
for $k=2,...,7$ on a logarithmic scale, we find that GALI$_2$
fluctuates around a non--zero value while all the other GALIs decay
to zero following power laws. This implies that the motion is that
of a quasiperiodic orbit that lies on a 2--dimensional (2D) torus.
In fact, for initial conditions (R2): $(x_j,y_j)$=(0.5,0.0),
$\forall$ $j \neq 11,12 $, perturbing two particles
$(x_{11},y_{11})=(0.65,0.0)$, $(x_{12},y_{12})=(0.55,0.0)$ for the
same parameters as in the R1 experiment, we detected regular motion
that lies on a 3D torus! This is demonstrated by the evolution of
the GALIs in Fig. \ref{1_2D_tor_real}, where not only GALI$_2$ but
also GALI$_3$ fluctuates around a non--zero value.

Such localized regular dynamics becomes evident in Fig.
\ref{1_2D_tor_real}c, where we exhibit the oscillations of the
$x_n$--coordinate of the $11^{th}$, $12^{th}$ and $13^{th}$ maps of
the system for the last $10^5$ iterations of the R2 experiment. As
seen in the figure, the motion is indeed quasiperiodic and confined
to the middle 3 maps, as all other degrees of freedom do not gain
any significant amount of energy. Similarly, we have observed that
exciting 3 central particles ($j=10,11,12$) gives a quasiperiodic
motion on 4D tori. This is reminiscent of the motion near discrete
breathers of Hamiltonian systems and suggests that the excitation of
each particle adds one extra frequency to the motion.

\textbf{B. Initial conditions localized in Fourier space}
\vspace{-0.5cm}

Finally, let us investigate the phenomenon of FPU recurrences in our
system of coupled standard maps. To do this, we first need to derive
the linear normal modes of the model, in order to choose initial
conditions that will excite only a small number of them. Thus,
keeping only the first term in the Taylor expansion of the
$sine$--function in (\ref{eq:CSM}), we obtain the following system
of equations: \vspace{-0.25cm}
\begin{equation}\label{li}
    x'=\mathcal{A} x,
\end{equation}
where \vspace{-0.25cm}
$x'=(x_{n+1}^{1},y_{n+1}^{1},...,x_{n+1}^{N},y_{n+1}^{N})^{T}$,
$x=(x_{n}^{1},y_{n}^{1},...,x_{n}^{N},y_{n}^{N})^{T}$ and
\begin{equation}
\mathcal{A}= \left[
\begin{matrix}
 K_1+\gamma & 1 & -\beta & 0 & \ldots & 0 & 0 & 0 & 0 & 0 \cr
 K_1+2\beta & 1 & -\beta & 0 & \ldots & 0 & 0 & 0 & 0 & 0 \cr
 -\beta  &   K_2+\gamma & 1 & -\beta & \ldots & 0 & 0 & 0 & 0 & 0 \cr
 -\beta  &   K_2+2\beta & 1 & -\beta & \ldots & 0 & 0 & 0 & 0 & 0 \cr
 \vdots & \vdots & \vdots & \vdots & & \vdots & \vdots & \vdots & \vdots \cr
 0 & 0 & 0 & 0& \ldots &  -\beta  &   K_{N-1}+\gamma & 1 & -\beta & 0 \cr
 0 & 0 & 0 & 0& \ldots &   -\beta  &  K_{N-1}+2\beta & 1 & -\beta & 0 \cr
 0 & 0 & 0 & 0& \ldots & 0 & 0 & -\beta &   K_N+\gamma & 1 \cr
 0 & 0 & 0 & 0& \ldots & 0 & 0 & -\beta &   K_N+2\beta & 1\cr
\end{matrix}
\right],
\end{equation}
with $\gamma = 1 + 2\beta$.

Using now well--known methods of linear algebra, we diagonalize the
matrix $\mathcal{A}$ writing $\mathcal{D}=\mathcal{P}^{-1}
\mathcal{A}\mathcal{P}$, where $\mathcal{P}$ is an invertible matrix
whose columns are the eigenvectors of the matrix $\mathcal{A}$ and
$\mathcal{D}$ is diagonal, $\mathcal{D}= diag
[\lambda_1,...,\lambda_{2N}]$, provided the eigenvalues $\lambda_i$
are real and discrete.

Our case, of course, involves oscillations about a stable
equilibrium point and hence the above system has $2N$ discrete
complex eigenvalues $\lambda_j=a_j + i b_j$, $\bar{\lambda}_j=a_j -
i b_j$  and eigenvectors $w_j=u_j + i v_j$, $\bar{w}_j=u_j - i v_j$,
with $j=1,2,...,N$. Thus, ${u_1,v_1,...,u_N,v_N}$ form a basis of
the space $\mathcal{R}^{2N}$ and the invertible matrix
$\mathcal{P}=[v_1,u_1,,...,v_N,u_N]$ leads us to the Jacobi normal
form:\vspace{-1.cm}
\begin{center}
\begin{eqnarray}
\mathcal{B}=\mathcal{P}^{-1} \mathcal{A} \mathcal{P}=diag \left[
{\begin{array}{*{20}c}
   a_j & -b_j  \\
   b_j &  a_j  \\
\end{array}} \right] \nonumber,
\end{eqnarray}
\end{center}
of a $2N \times 2N$ matric $\mathcal{B}$ with $2 \times 2$ blocks
along its diagonal. In this way, using the transformation
$z=\mathcal{P}^{-1} x$, we can reduce the initial problem to a
system of uncoupled equations, whose evolution is described by the
equations:
\begin{equation}\label{eq:unc}
z'=\mathcal{P}^{-1} \mathcal{A} \mathcal{P} z,
\end{equation}
where
$z'=(l_{n+1}^{1},k_{n+1}^{1},...,l_{n+1}^{N},k_{n+1}^{N})^{T}$,
$z=(l_{n}^{1},k_{n}^{1},...,l_{n}^{N},k_{n}^{N})^{T}$ and $z$
represents the linear normal modes of the map. Thus, in order to
excite a continuation of one or more of these modes, we extract the
appropriate $x$--initial condition by the transformation
$x=\mathcal{P}z$. The evolution in the uncoupled coordinates is
given by the transformation: \vspace{-0.5cm}
\begin{eqnarray}
  l_{n+1}^{j} &=& a_i l_{n}^{j} - b_i k_{n}^{j} \\ \nonumber
  k_{n+1}^{j} &=& b_i l_{n}^{j} + a_i k_{n}^{j}
\end{eqnarray}
where each pair of $(l^j,k^j)$ corresponds to a normal mode of the
system. Since the map is area--preserving $a_j^2 +b_j^2=1$ and we
may define the quantity $E_n^j=(l_n^j)^2+(k_n^j)^2$ as the energy of
each mode, which is preserved under the evolution of the linear map,
i.e. $E_{n+1}^j=E_n^j$.

\begin{figure*}[t]
  \includegraphics[width=8.5cm]{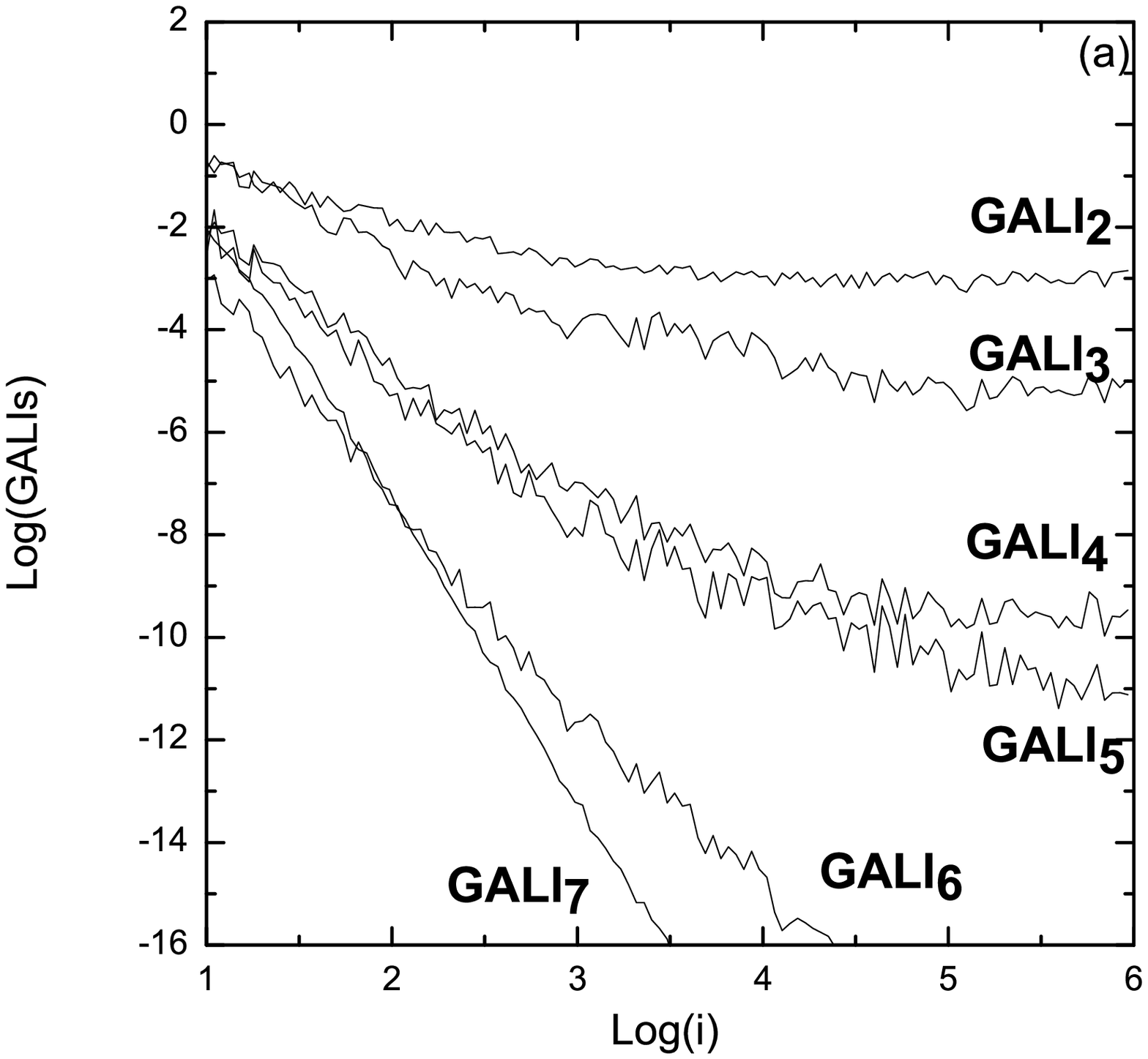}\hspace{-1.5cm}
  \includegraphics[width=8.5cm]{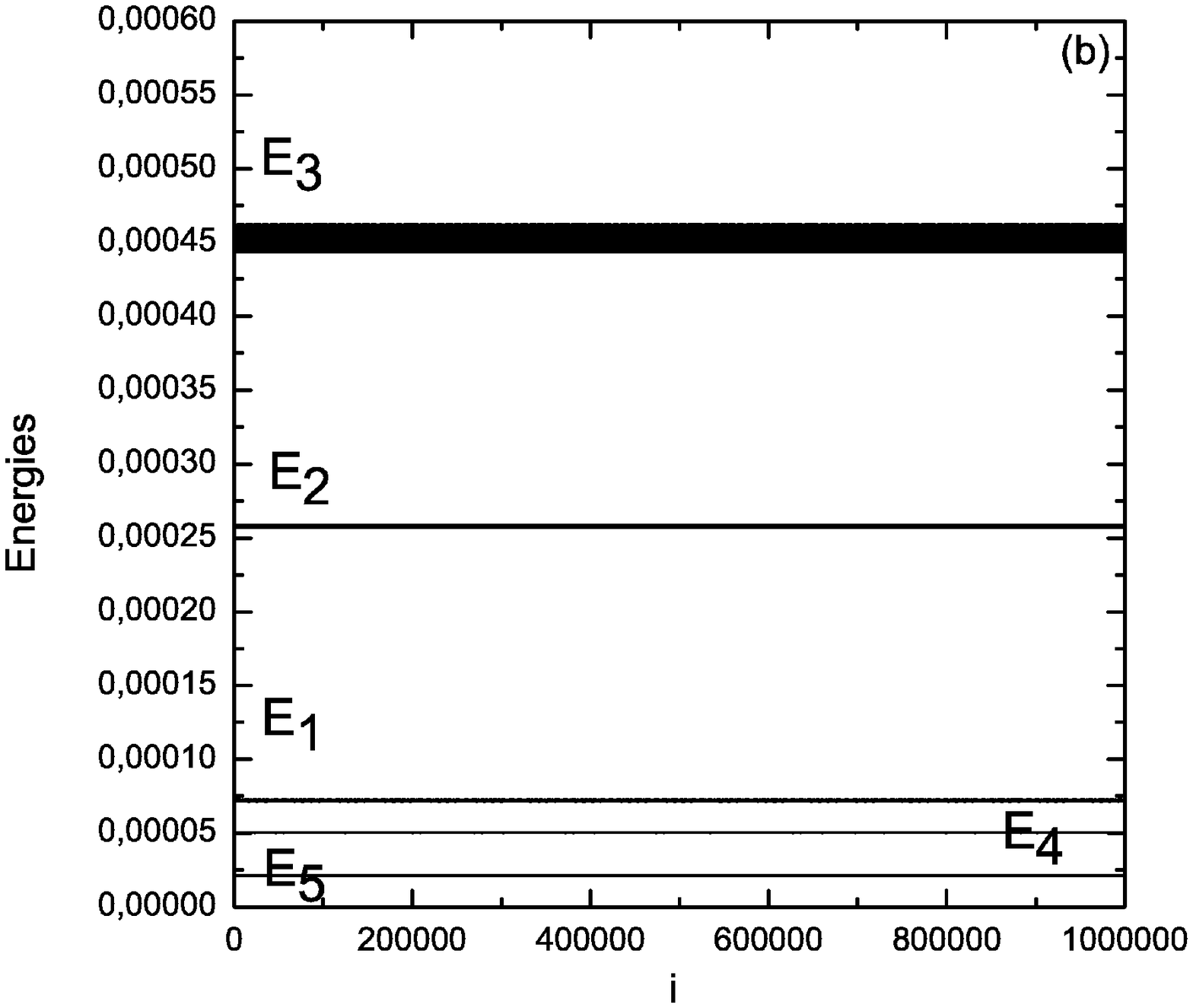}
  \caption{(a) GALIs for 5 coupled standard maps with $K_1=-2.30,K_2=-2.35,K_3=-2.40,K_4=-2.45,K_5=-2.50)$, $B=0.00001$
   and initial conditions that excite 5 different normal modes. Note that the GALI$_k$ for $k=2,3,4,5$ fluctuate around
   a non--zero value implying a regular motion that lies on a 5D torus. (b) The energies for the corresponding normal modes.}\vspace{0.5cm}
   \label{10D_sm_5D_torus_modes}
\end{figure*}

Thus, to study recurrences, we start with the example of $N=5$
coupled standard maps, choose a `small' coupling parameter
$\beta=0.00001$ and excite 5 different modes. As a result, we are
able to detect a quasiperiodic orbit that lies on a 5D torus,
indicated by all the GALI$_k$, $k=1,...,5$, approaching a constant,
as shown in Fig. \ref{10D_sm_5D_torus_modes}. This is verified in
Fig. \ref{10D_sm_5D_torus_modes}b, where we plot the corresponding
energy modes of the coupled system and find that they are nearly
invariant, as they would have been in the uncoupled case.
%=============================================
%=============================================
%=============================================
\begin{figure*}[t]
  \includegraphics[width=8.5cm]{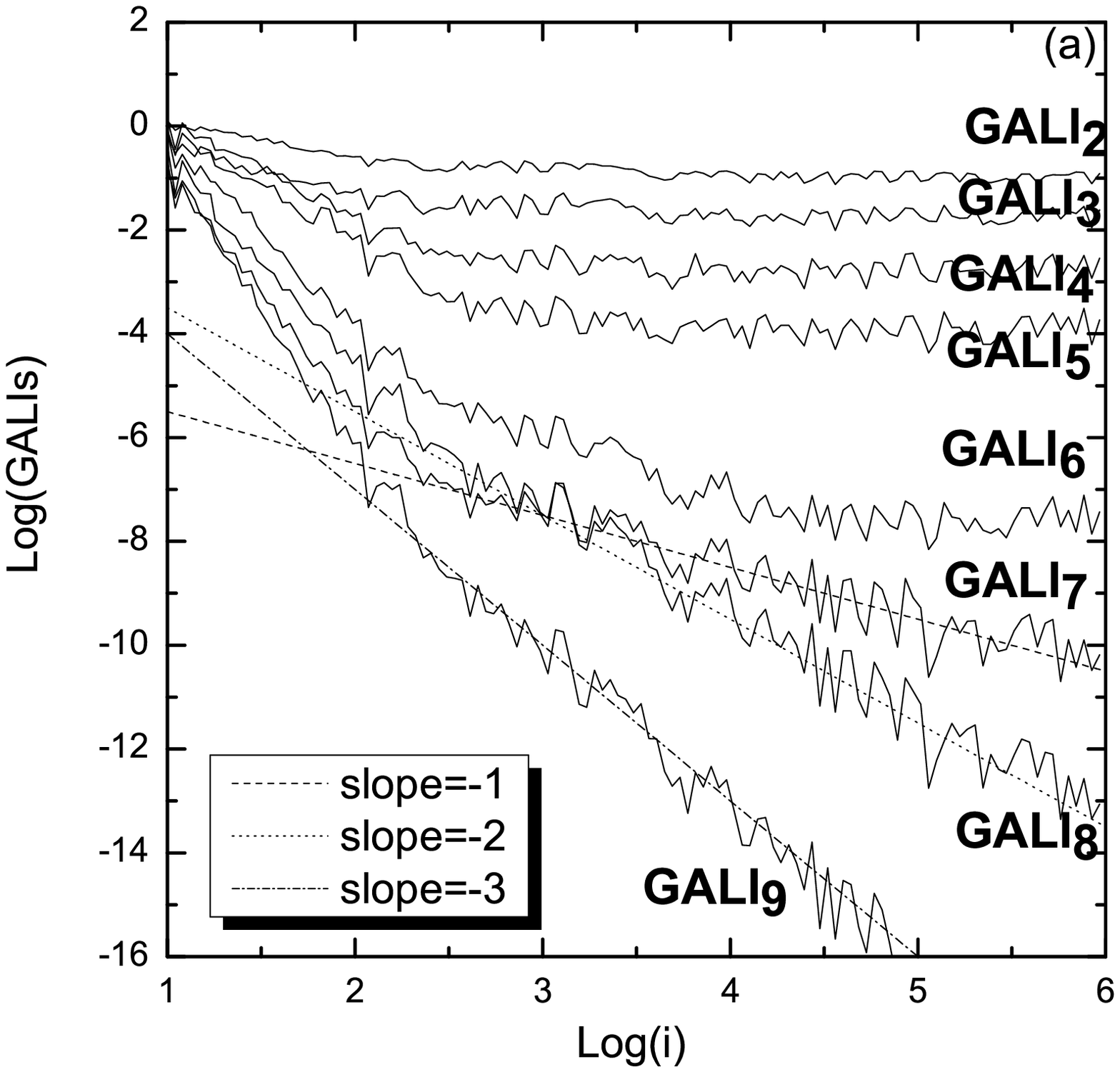}\hspace{-1.5cm}
  \includegraphics[width=8.5cm]{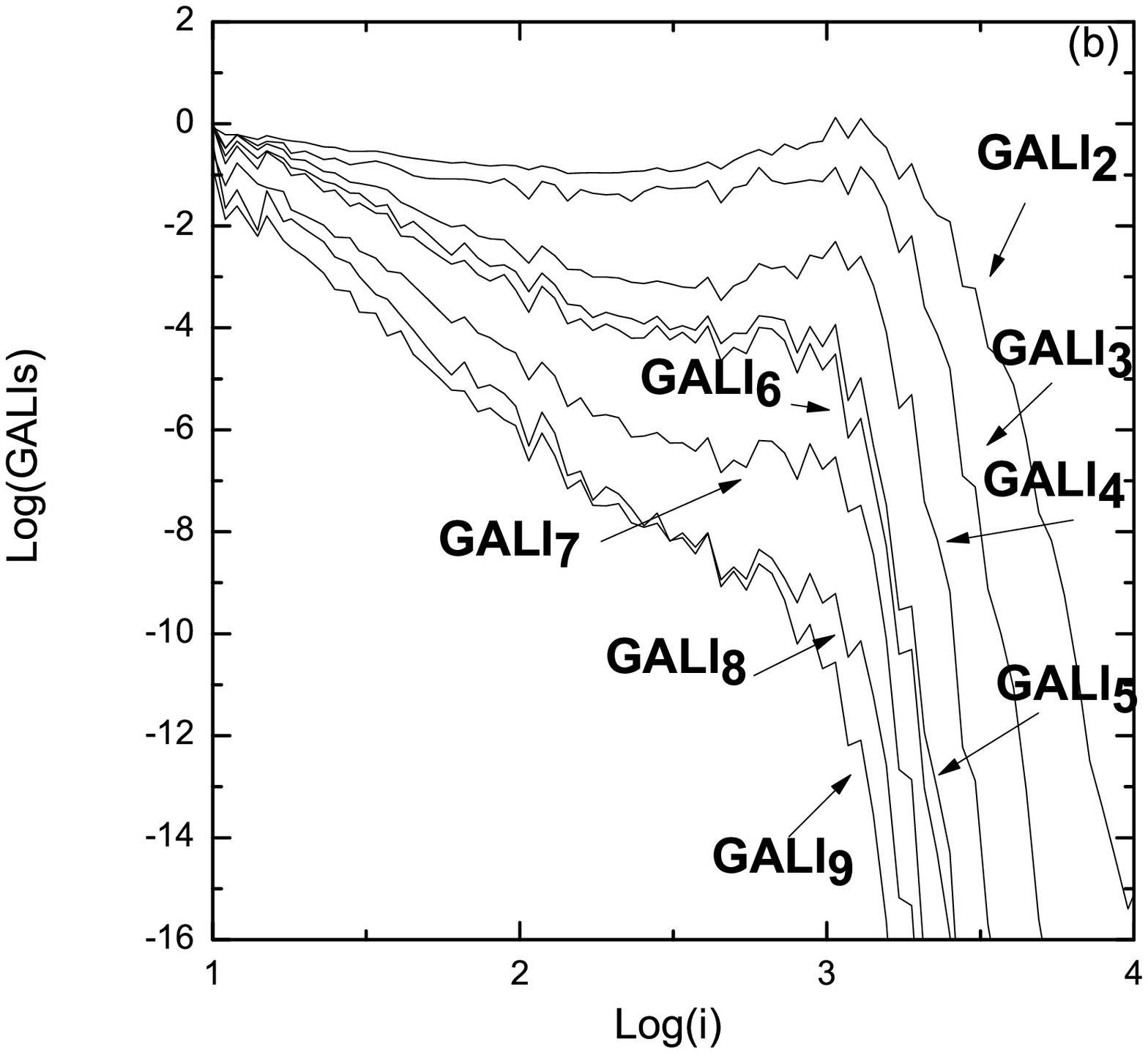}
  \caption{(a) Exciting only one mode in a case of 20 coupled standard maps, with $\beta=0.00001$ and $K_j$ in triplets of
  $-1.35, -1.45, -1.55$, the GALI$_k$, for $k=2,...,6$ asymptotically become constant, implying that the motion lies on a 6D torus.
  (b) An orbit with the same initial condition and $K_j$ as in (a) but with a larger coupling parameter, ($\beta=0.01$),
  becomes chaotic, as all the GALIs are seen to decay exponentially to zero.}\vspace{0.5cm}
  \label{20_sm_6D_torus_chaos}
\end{figure*}
%=============================================

Next, we increase the number of coupled maps from 5 to 20, keeping
$\beta=0.00001$ and choosing different $K_{j}$ (in triplets of
-1.35, -1.45, -1.55), with $j=1,...,20$. Exciting now only
\textit{one normal mode} (the 21$^{st}$), much like the original FPU
case, we calculate the GALIs again and find that the orbit is again
quasiperiodic, lying on a 6--dimensional torus, as evidenced by the
fact that GALI$_k$ for $k=2,...,6$ fluctuate around non--zero
values, see Fig. \ref{20_sm_6D_torus_chaos}a. This is, of course, a
low--dimensional torus, since generically one expects regular orbits
of this system to lie on $N=20$--dimensional tori! From the
calculation of the normal mode energies in this experiment, we
observe that the initial energy is now shared by 6 modes and
executes recurrences, for as long as we have integrated the
equations of motion. The energies of these 6 modes are not shown
here since their values are too close to each other to be
distinguished in a graphical representation.

Finally, as our last experiment, we increase the coupling parameter
to $\beta=0.01$, for the same choice of initial conditions and $K_j$
values as in Fig. \ref{20_sm_6D_torus_chaos}a. The result is that
the influence of the coupling is now quite strong and leads to the
breakdown of recurrences and the onset of chaotic behavior. This is
clearly depicted in Fig. \ref{20_sm_6D_torus_chaos}b, where the
GALI$_k$s of this case exhibit exponential decay, implying the
chaotic nature of this orbit.

\vspace{-0.5cm}
%=============================================
\section{Conclusions}\vspace{-0.75cm}
\label{concl}

Localized oscillations in one--dimensional nonlinear Hamiltonian
lattices constitute, for the last 15 years, one of the most active
areas of research in Mathematical Physics. Of primary interest has
been the discovery of certain exact periodic solutions in such
lattices, called discrete breathers, which are exponentially
localized \textit{in real space}. It is known that when these
solutions are stable, there are regions around them in phase space,
where orbits oscillate quasiperiodically for very long times, even
though the presence of linear dispersion is expected to lead them to
delocalization away from the discrete breather. How long do they
stay, however, in that vicinity and what is the dimensionality of
the tori on which they lie? Furthermore, how large are these regular
regions and at what initial conditions (or parameter values) does
chaotic behavior begin to arise?

Similar questions can also be posed about another form of
localization \textit{in Fourier space}, which has been very recently
proposed as an explanation of the famous recurrence phenomena known
to characterize the dynamics near linear normal modes of nonlinear
lattices, since the early 1950s. Clearly, the interest in such
localization lies in the (in)ability of Hamiltonian lattices to
exhibit energy propagation and equipartition as expected from the
theory of statistical mechanics.

To answer such questions, we have applied in this paper a new method
that our group has recently developed, which is able to distinguish
order from chaos, identify the dimensionality of tori and predict
slow diffusion in thin chaotic layers, faster and more reliably than
other methods (see also our recent publication \cite{sk:7}). This is
accomplished by computing what we call the \textit{GALI indices} of
the orbits under study, for which we have obtained analytical
formulas that are in excellent agreement with our numerical
calculations.

We have thus considered two examples of high dimensional systems:
One Hamiltonian lattice described by 31 Newton's second order
differential equations and one system of 20 coupled area--preserving
pairs of difference equations mapping the plane onto itself. In each
of these models, we have been able to demonstrate the usefulness of
the GALI indices in analyzing the properties of regular motion
localized in real and Fourier space. Furthermore, we have shown that
our indices can accurately determine cases where the motion becomes
delocalized and starts to diffuse chaotically in phase space, long
before this is observed in the actual oscillations.

We wish to emphasize the superiority of our approach to the
traditional calculation of \textit{Lyapunov exponents}, still used
by many researchers to distinguish regular from chaotic motion. As
we have shown, Lyapunov exponents can be misleading, as they often
tend to decrease in time giving the impression that the motion is
quasiperiodic, while it is in fact weakly chaotic. The GALI indices,
on the other hand, are computationally more efficient, as they
concentrate on the \textit{tangent space} of the dynamics and
explore the \textit{linear (in)dependence} of $k$ unit deviation
vectors about an orbit representing the volume of their associated
parallelepiped. If, for $k=1,2...,d$, that volume remains constant
and decays for $k>d$ by well--defined power laws, the motion is
quasiperiodic on a $d$--dimensional torus, while if the volume
decreases exponentially for all $k$ the motion is chaotic. We,
therefore, hope that the virtues of the GALI method emphasized in
this paper, will encourage researchers studying the dynamics of
multi--dimensional conservative systems to use them in order to
study many other physically interesting applications.

\vspace{-0.5cm}
%=============================================
\section{Acknowledgements}\vspace{-0.75cm}
The authors wish to thank Dr. Haris Skokos and Dr.Christos
Antonopoulos for many useful discussions on the topics of this
paper. T.~Manos was partially supported by the ``Karatheodory"
graduate student fellowship No B395 of the University of Patras, the
program ``Pythagoras II" and the Marie Curie fellowship No
HPMT-CT-2001-00338. H. Christodoulidi was supported by the State
Scholarships Foundation of Greece.

\vspace{-0.5cm}
%=============================================

\end{document}